\documentclass[sigconf, balance=false]{acmart}

\usepackage[normalem]{ulem}

\AtBeginDocument{%
  \providecommand\BibTeX{{%
    \normalfont B\kern-0.5em{\scshape i\kern-0.25em b}\kern-0.8em\TeX}}}

\usepackage{xurl}
\usepackage{amsmath,amsfonts,chemarrow,balance, graphicx, subcaption, url, multirow, graphics, booktabs, colortbl, algpseudocode, xcolor}
\newcommand{\BfPara}[1]{\vspace{1mm}{\noindent\bf#1.}\xspace\xspace}
\usepackage[linesnumbered,ruled]{algorithm2e}

\usepackage{tikz}

\usepackage{hyperref}
\usepackage[normalem]{ulem}

\usepackage[linesnumbered,ruled]{algorithm2e}
\usepackage{tikz}
\usepackage{hyperref}
\usepackage[normalem]{ulem}
\usepackage{tcolorbox}

\usepackage{booktabs}

\usepackage[font=footnotesize,labelformat=simple]{subcaption}

\usepackage[framemethod=TikZ]{mdframed}
\definecolor{rubcolor}{HTML}{7E8995}
\global\mdfdefinestyle{insightstyle}{%
backgroundcolor=rubcolor!3,
outerlinewidth=1pt,innerlinewidth=0pt,
outerlinecolor=rubcolor,roundcorner=5pt
}
\newmdenv[roundcorner=10pt, frametitle=Key Takeaways, linecolor=rubcolor]{insightbox}

\usepackage{pifont}
\newcommand{\circleone}{\ding{202}\xspace}
\newcommand{\circletwo}{\ding{203}\xspace}
\newcommand{\circlethree}{\ding{204}\xspace}

\newcommand{\spacesep}{\ensuremath{\quad}}

\hypersetup{
    pdfpagemode=pagewidth,
    plainpages=false,
    colorlinks,
    urlcolor=blue!70!black,
    linkcolor=red!70!black,
    citecolor=green!70!black,
}

\let\oldautoref\autoref
\renewcommand{\autoref}[1]{\textcolor{red!70!black}
{\oldautoref{#1}}}

\usepackage{paralist}

\renewcommand\footnotetextcopyrightpermission[1]{} % removes footnote with conference information in first column
\usepackage{popets}

\setcopyright{popets}
\copyrightyear{2025}

\acmYear{2025}
\acmVolume{2025 }
\acmNumber{v 1.0}
\acmDOI{XXXXXXX.XXXXXXX}
\acmISBN{}
\acmConference{Under Review at WWW}
\author { \large {Bhupendra Acharya\textsuperscript{\textdagger} Dario Lazzaro\textsuperscript{\$} Antonio Emanuele Cinà\textsuperscript{\$}  Thorsten Holz\textsuperscript{\textdagger}} \\ {\textsuperscript{\textdagger}CISPA Helmholtz for Information Security, \textsuperscript{\$}Università di Genova} \\
\texttt{bhupendra.acharya@cispa.de} \spacesep \texttt{dario.lazzaro@edu.unige.it} \spacesep \texttt{antonio.cina@unige.it} \spacesep\texttt{holz@cispa.de}}

\begin{document}
\title[Exploring Donation-based Abuses in Social Media Platforms]{\huge \bf \emph{Pirates of Charity}: Exploring Donation-based Abuses in Social Media Platforms}

\renewcommand{\shortauthors}{Acharya et al.}

\begin{abstract}

With the widespread use of social media, organizations, and individuals use these platforms to raise funds and support causes. Unfortunately, this has led to the rise of scammers in soliciting fraudulent donations. In this study, we conduct a large-scale analysis of donation-based scams on social media platforms. More specifically, we studied profile creation and scam operation fraudulent donation solicitation on X, Instagram, Facebook, YouTube, and Telegram. By collecting data from 151,966 accounts and their 3,053,333 posts related to donations between March 2024 and May 2024, we identified 832 scammers using various techniques to deceive users into making fraudulent donations. Analyzing the fraud communication channels such as phone number, email, and external URL linked, we show that these scamming accounts perform various fraudulent donation schemes, including classic abuse such as fake fundraising website setup, crowdsourcing fundraising, and asking users to communicate via email, phone, and pay via various payment methods. Through collaboration with industry partners PayPal and cryptocurrency abuse database Chainabuse, we further validated the scams and measured the financial losses on these platforms. Our study highlights significant weaknesses in social media platforms' ability to protect users from fraudulent donations. Additionally, we recommended social media platforms, and financial services for taking proactive steps to block these fraudulent activities. Our study provides a foundation for the security community and researchers to automate detecting and mitigating fraudulent donation solicitation on social media platforms. 

\end{abstract}
%\keywords{Donation abuse, fraudsters, social media, web security}

\maketitle

\section{Introduction}
\label{sec:introduction}
Recently, there has been an increase in fraudsters using social engineering tactics to trick people into donating to fake charities or causes~\cite{donationfraudefFTC,ftcscamcharityscam,fbicharitydisasterfraud}. These tricks often include playing on sympathy and asking for a donation. Traditionally, fraudsters perform such attacks via the setup of fake donation websites~\cite{fastcompanycharityscam}, impersonation via phone calls~\cite{fakecharitycalls}, sending an e-mail or text asking to donate to a charity or cause~\cite{fakeEmailTextCharity}, and sending a return letter envelope asking a cheque to send via mail~\cite{fakecharitydonationemailcalls}. As there has been a rise in social media users sharing, organizing, and participating in charity-related causes, this has simultaneously led to fraudsters shifting to conducting various donation scams on these platforms~\cite{twitterfakeprofileCharityScam,howtospotCharityScamBitdefender}. 
Donation fraud, which is also commonly known as charity scam, is where scammers solicit money from individuals in the pretense of a charitable cause, disaster relief, or other seemingly legitimate reasons~\cite{donationfraudefFTC,britannicacharityfraud}. 
These fraudulent activities can occur through various means, including fake websites, emails, social media posts, and crowdfunding platforms~\cite{fakecharityemailscaf,ftcscamcharityscam,universityofReginacharityscam,fastcompanycharityscam}. The scammers deceive donors by pretending to represent real charities or by creating fictitious causes, often using emotional appeals to make urgent donations. Once the money is donated, it is typically diverted for the scammer's personal use, and the intended cause or individuals in need receive no benefit~\cite{ftcscamcharityscam}.
\begin{figure*}[tb]
    \centering
    \subcaptionbox{\centering Facebook Donation Scam Profile Page\label{fig:ts_a}}{\includegraphics[width=0.227\textwidth]{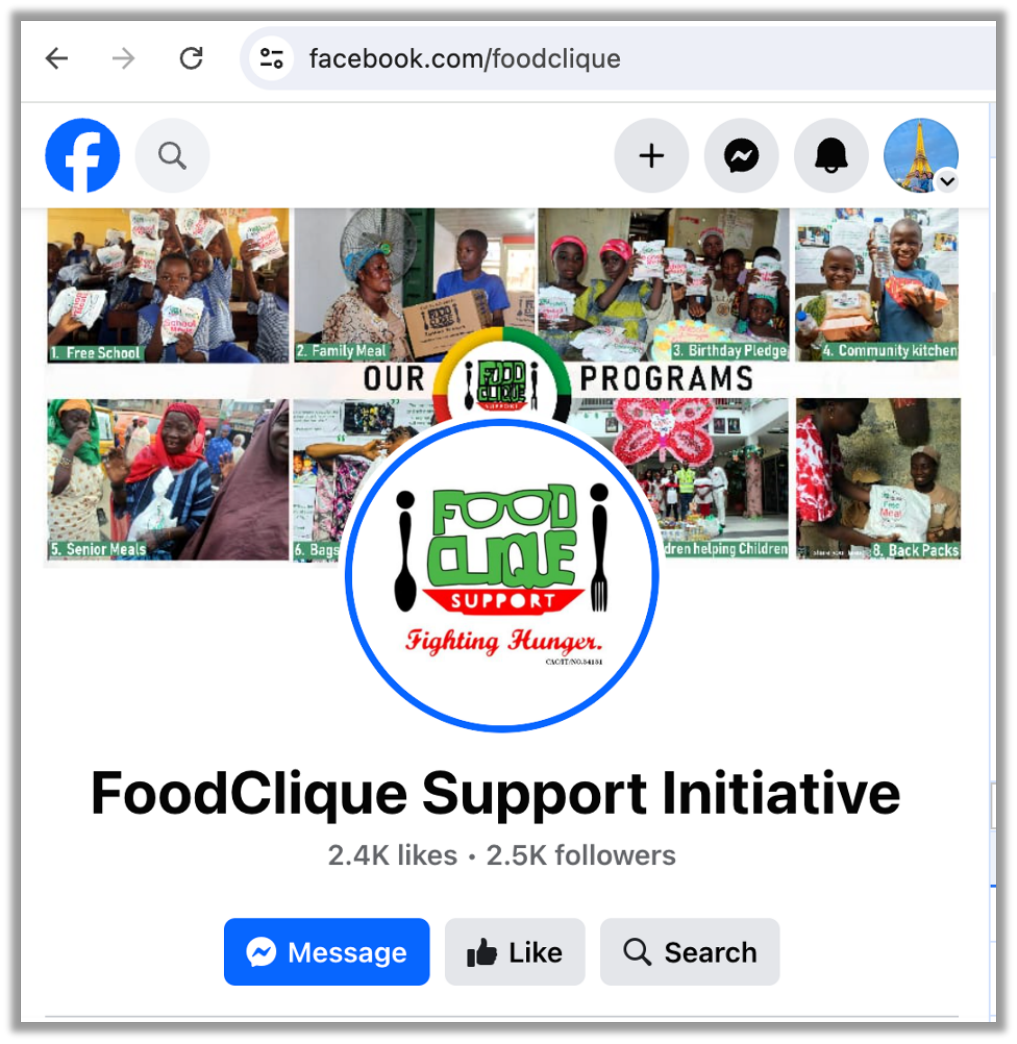}}\hfill
    \subcaptionbox{\centering Instagram Donation Scam Profile Page\label{fig:ts_b}}{\includegraphics[width=0.247\textwidth]{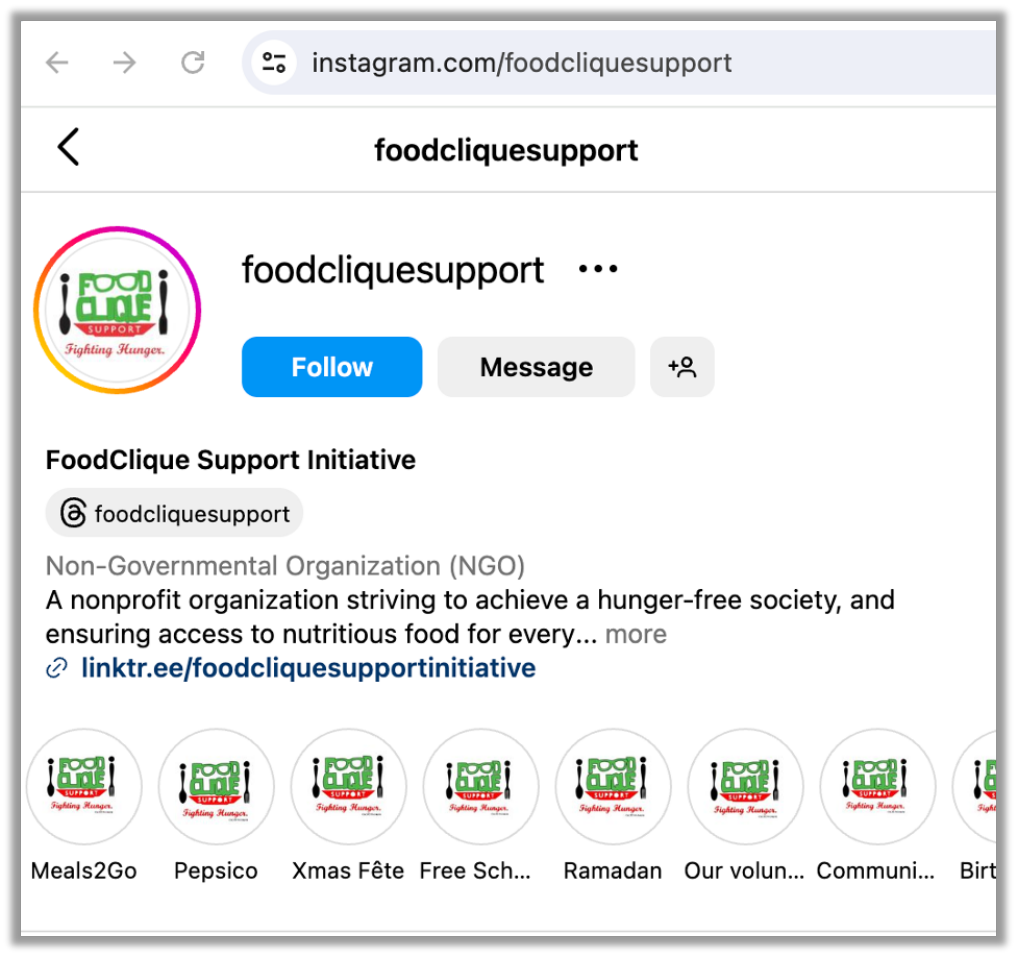}}\hfill
    \subcaptionbox{\centering Website Affiliated to Scamming Profile\label{fig:ts_c}}{\includegraphics[width=0.206\textwidth]{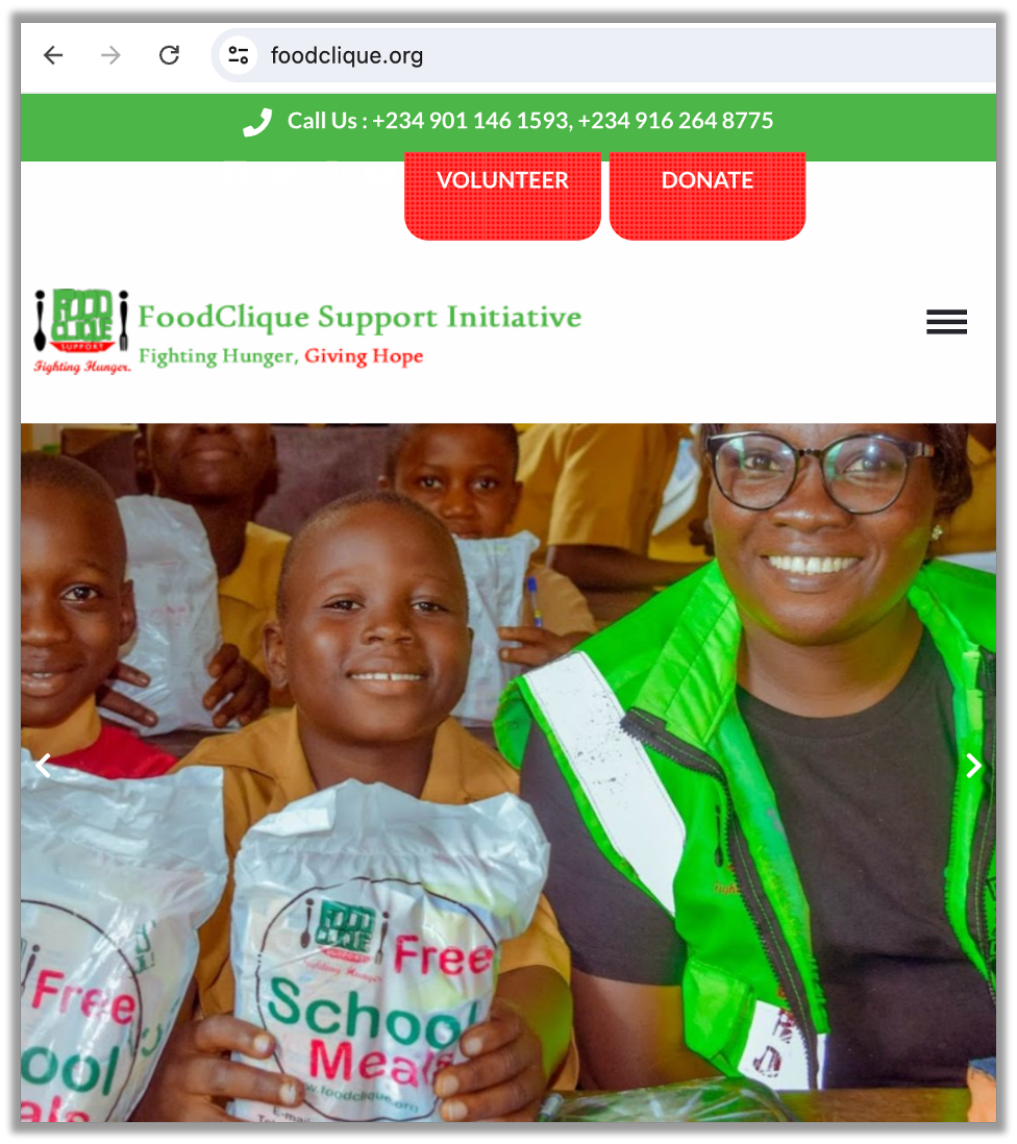}}\hfill
    \subcaptionbox{\centering VirusTotal Anti-Phishing Engine Fraud Analysis\label{fig:ts_d}}{\includegraphics[width=0.264\textwidth]{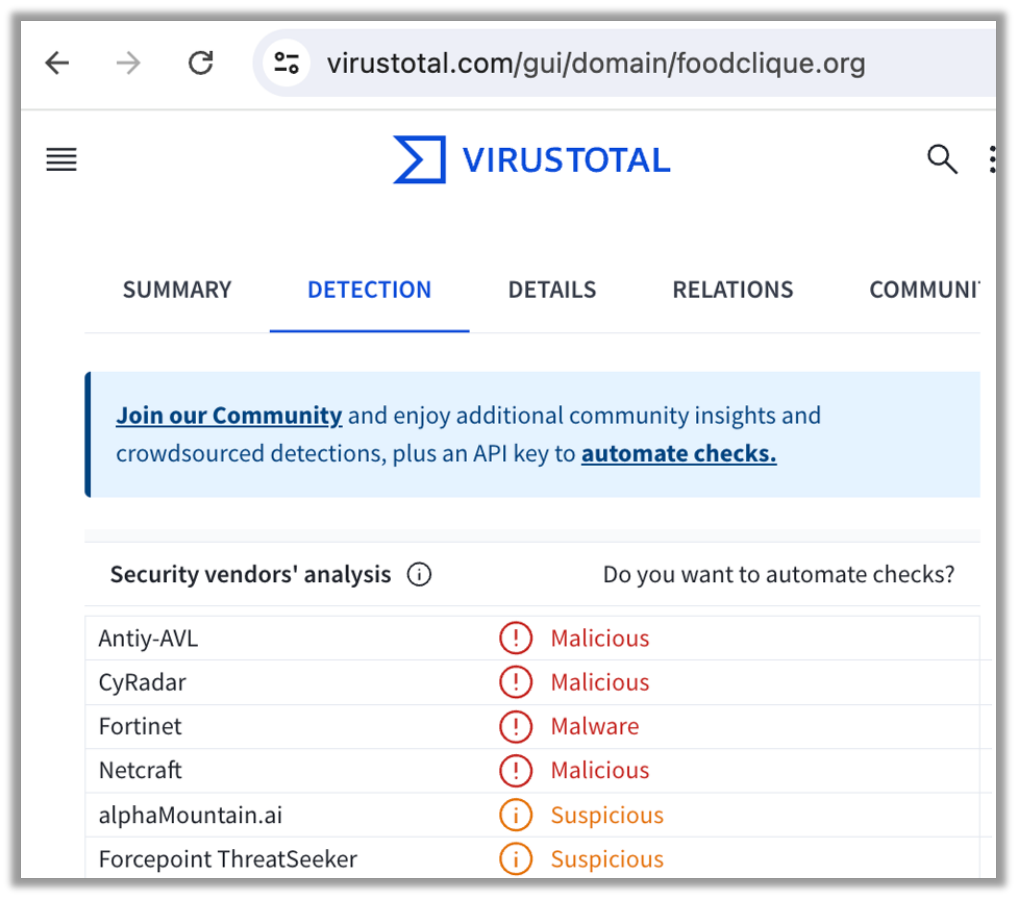}}
    \caption{Examples of Scamming Donation Support Request: The first two images \autoref{fig:ts_a}, \autoref{fig:ts_b} show the associated social media profile of the scamming donation on Facebook and Instagram social media platforms. The third image \autoref{fig:ts_c} shows the associated external website asking for a donation to support and the last screenshot \autoref{fig:ts_d} shows the risk engine evaluation from multiple anti-phishing engines (\emph{Antiy-AVL, CyRadar, Fortinet, Netcraft, AlphaMountain.ai} and \emph{Forcepoint ThreatSeeker}) indicating that the website is malicious or suspicious. The social media profiles can appear genuine, making it difficult to recognize the scam at first glance.}
    \label{fig:scamming_donation_pages}
\end{figure*}
Over the years, social media users have steadily grown and are projected to reach 5 billion by 2025~\cite{backlinkosocialmediausers}. 
Social media is popular among legitimate organizations and individuals to request donations for various causes~\cite{socialmediaStatCharity}. 
It provides building networks and easy sharing for users and charitable organizations through posts, tags, and direct message communications~\cite{socialmediaUserForBoost}. 
Unfortunately, as social media adoption for donations has increased, fraudsters have also shifted towards social media-based donation scams. These scams include but are not limited to impersonating profiles of well-known organizations, individuals, or family members. Scammers often try reaching out by sending thank-you notes via direct messages, tagging posts for donations that users never made, or sending a friend or network requests to further establish a connection in the act of performing donation-based scams~\cite{socialmediaReceivedMessage,attSocialMediaCharityScam}. 

According to the FTC, social media-based scams are on the rise, with more than  $2.7$ billion in losses from 2021 to 2023~\cite{ftcsocialmediaattack}. Social media offers easy account creation compared to launching web domains, which often requires going through hosting websites and content. Various donation scams are increasing, with fraudsters posing as reputable organizations and soliciting contributions~\cite{fbicharitydisasterfraud,dirsdirtydozen, cnbccharityfraud}. Scammers performing donation-based abuse in social media are ever rising~\cite{nitsocialmediafraud, scamstatistics,euronews2023}, and with the rise of AI tools and content creation scammers are trending to abuse social media higher than before~\cite{howtogeek}. With the wide adoption of cryptocurrency globally, scammers are also shifting towards requesting donations via cryptocurrency~\cite{cyberscoops2022,Abnormal2023,ZKENews} and using crypto drainers as part of the fraud. These crypto drainers trick victims into connecting through fake web wallet browsers, stealing their private key phrases, and ultimately draining the total funds from their wallets~\cite{cryptodrainercheckpoint2023,netcraft2024}. In appendix~\autoref{fig:scamming_donation_pages}, we display an example of fraudulent donation soliciting on multiple platforms. Despite fraudulent donations being rampant on social media, there still lacks an end-to-end life cycle study of scammers' behavior, operation, and financial impact. 

In this work, we address the research gap in donation-based abuses by conducting a study across five social media platforms. We assess profile creation, user engagement, and the external communication channels that scammers use to solicit contact and payments for fraudulent donation scams. Specifically, we conduct the first large-scale study of donation-based abuses on ~\emph{X}, ~\emph{Instagram}, ~\emph{Telegram}, ~\emph{YouTube}, and ~\emph{Facebook}. Using donation-related search contexts, we collected data from 150K social media users and 3M posts. By analyzing the scammers' profile metadata and posts, including fraudulent emails, phone numbers, and URLs, we identified 832 scammers conducting fraudulent donation solicitations across these platforms. Additionally, our network analysis on these scamming accounts uncovers an additional 1K accounts linking to 11 platforms beyond their originating platforms. Furthermore, we provide an in-depth analysis of the scamming profiles' account creation, engagement posts, and techniques used to lure victims into fraudulent donations. Our findings show that social media platforms are not effectively blocking fraudulent accounts or protecting users against such abuses. Finally, we offer recommendations for proactive blocking and mitigation of these fraudulent activities for various platforms and payment processors. 

~\textbf{}

\noindent
\textbf{Contributions.} Our key contributions are as follows:

~\textbf{}

\begin{itemize}
    \item ~\textbf{Fraudulent Donation Solicitation Measurement.} We conduct the first large-scale study of fraudsters soliciting donations across multiple social media platforms. Our approach uncovers scam accounts and their interconnected operations extending beyond their original platforms. 

    \textbf{} 
    
    \item \textbf{Fraudulent Payment Detection.} We identify fraudulent payment profiles and channels used by scammers to collect payments for fake donations. This enables tracking of financial losses and provides a blueprint for financial services to implement proactive solutions for detecting payment-related fraud.
\end{itemize} 
 
\textbf{Ethical Concerns and Data Disclosure.} Our research did not involve interaction with any human subjects, including scammers. We collected public data from social media profiles using API queries. Additionally, we disclosed our findings to all five social media platforms:~\emph{X}, ~\emph{Instagram}, ~\emph{Facebook}, ~\emph{YouTube}, and ~\emph{Telegram}. For payment profiles linked to scamming accounts, we collaborated with ~\emph{PayPal} and the cryptocurrency abuse database ~\emph{Chainabuse}, both of which provided positive feedback and scam validation. We also shared email addresses, phone numbers, crowdfunding URLs, and survey forms associated with these scamming profiles with their respective service providers. ~\emph{PayPal} confirmed that the flagged accounts were involved in various nefarious activities. ~\emph{Chainabuse’s} evaluation of cryptocurrency addresses revealed the scale of these attacks and associated financial losses. In summary, our work received several positive acknowledgments and validation of the abuses caused by fraudulent social media profiles soliciting donations. We provide our research code in a GitHub repository~\cite{scamdonationgitHub} to foster future research. However, data related to scammers will be only shared with the researcher upon request to prevent potential retribution attacks.

\section{Related Work}
\label{sec:related_work}
In this paper, we perform a holistic study of scammers performing donation-based abuses across five social media platforms. To the best of our knowledge, we are the first to perform a large-scale analysis of donation-based abuses orchestrated by fraudsters on multiple platforms. Given the extensive research on scams and abuses over the past two decades, in this section, we focus on how our work diverges from previous studies and highlight the novelty of our approach in validating donation-based abuses.

% \textbf{}

\BfPara{Domains: Abuses, Scams, and Attacks Study} The use of web domains for distributing scams, and attacks remains a potent channel for abusers and has been widely researched over the last decade. These include studies such as traditional phishing attacks~\cite{oest2019phishfarm,oest2020sunrise,acharya2021phishprint,peng2019,peng2021,subramani22}, Technical Support Scams~\cite{liu2023understanding,miramirkhani2016dial}, and beyond such as Squatting-based attacks~\cite{liu2016towards,nikiforakis2014soundsquatting,nikiforakis2013bitsquatting,agten2015seven,wang2006strider,szurdi2014long}, and Malvertisement~\cite{vadrevu2019you,zarras2014dark,SrinivasanKMANA18}. For instance, in PhishFarm~\cite{oest2019phishfarm}, the author studied how malicious actors evade the anti-phishing engines in distributing various forms of scams and abuses in web domains. Agten et al.~\cite{agten2015seven} studied squatting-based attacks that malicious actors perform via registering the squatting domains. With the rise of the adoption of digital currency over recent years, online frauds and attacks related to cryptocurrency scams are found ever rising, and tracking this fraud has caught the interest of security communities~\cite{xigao2023doublenothing,9493255,phillips2020tracing,hong2021mobilescams}.

% \textbf{}

\BfPara{Social Media: Abuses, Scams, and Attacks Study}
With the rise of abuses, scams, and attacks in social media platforms, social media has been a platform of interest to measure the prevalence of abuses among security communities and researchers. These studies explored various categories of social media scams including but not limited to Technical Support Scams~\cite{acharya2024conning}, Comment Scams~\cite{li2024commentscams}, Cryptocurrency Abuses~\cite{mirtaheri2021identifying}, Fake Profiles~\cite{khaled2018detecting,mink2022deepphish} and Impersonation Attacks~\cite{acharyaImpersonation24} revealing the widespread nature of these issues on social media. Abusers continuously develop new attacks, making detecting malicious profiles based on publicly available data has become increasingly challenging for the security community and researchers. For example: in HoneyTweet~\cite{acharya2024conning}, Acharya et al. studied creating baiting tweets to lure scammers into an interaction with the posted tweets and performed an interaction with scammers to identify the modus operandi. The author also continued studying the variety of attacks that abusers perform as part of impersonating brands in the top 10K brands in multiple social media~\cite{acharyaImpersonation24}. The most relevant work to us in areas of YouTube-based comments was studied by Li et al.~\cite{lilike}, which analyzed scam campaigns and evasion techniques that scammers distributed as part of interacting comments on YouTube.

% \textbf{}

\BfPara{Donation Abuse Study} In areas of donation-based study, some of the prior work that are most relatable are from ~\cite{whitty2020there, albanese2005fraud, gillespie2020tackling, wood2022scams}. Whitty et al.~\cite{whitty2020there} examined the psychological profiles of cyber scam victims and the types of scams associated with these profiles. Among these scams, one of the scams studied on charity scams involving fake profiles and organizations that deceive victims into donating to fraudulent causes. Korsell et al.~\cite{albanese2005fraud} explored a taxonomy of fraud prevalent in 2020, highlighting the rise of charity and consumer scams. Similarly, Wood et al.~\cite {wood2022scams} studied the various scams that were found emergent during COVID-19 and touched upon charity-based scams that were rampant during COVID-19. However, neither of these studies provided an in-depth analysis of how donation-based scams are propagated via social media platforms or the lifecycle of these scams as conducted by malicious actors. 

% \textbf{}

\BfPara{Novelty} The prior work on social media has predominantly focused on other forms of attacks. Addressing this gap, our research performs an in-depth analysis of donation-based abuses on social media and their validation as scams. We leverage a straightforward methodology backed by LLMs and security risk engines well suited for identifying fraudulent profiles soliciting donations. The novelty of our work lies in identifying large-scale donation-based abuses across multiple platforms beyond the originating social media platforms and validating these scams through the association of fraudsters' payment profiles.

\section{Evaluation Setup and Data Filtration}
\label{sec:evaluation}

\begin{figure}[tb]
\centering
\includegraphics[width=.42\textwidth]{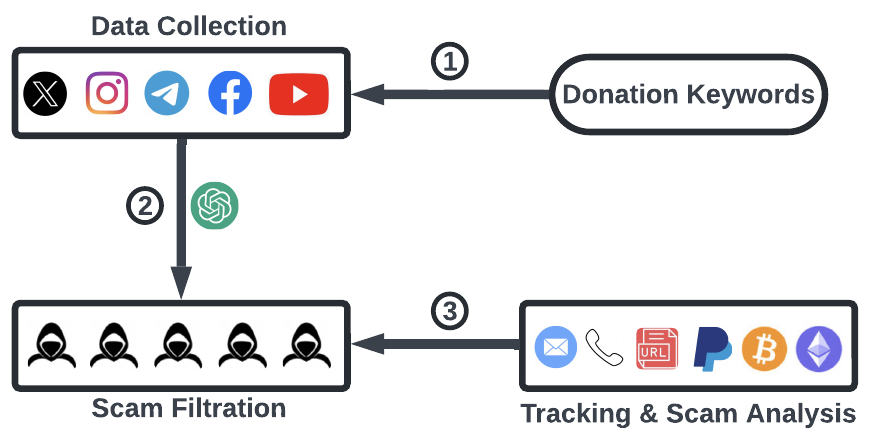}\hfill
\caption{\emph{Evaluation Setup Design}: An overview of our system, which consists of mainly three components: (i) \emph{Data Collection} which performs automated donation-based keyword searches in five social media platforms, (ii) \emph{Scam Filtration} which performs data filtration associated to donation soliciting fraudulent accounts, and (iii) \emph{Tracking and Scam Analysis} which provides an evaluation of scammer's modes of operation and techniques.}
\label{fig:sys_design}
\end{figure}

In this section, we detail our evaluation setup for identifying abusive social media profiles, particularly those soliciting fraudulent donations. We start by collecting data, including associated posts, from various social media platforms. This data is then filtered to focus on fraudulent donation solicitations, enabling a deeper analysis of scam operations. As shown in ~\autoref{fig:sys_design}, our measurement setup consists of three main components: \circleone, which gathers data from various social media platforms using donation-related keywords; \circletwo, which filters the data to pinpoint profiles involved in donation scams; and \circlethree, which tracks the scammers' methods of operation. We provide details for each component as below.

\subsection{Raw Dataset Aggregation} 
In order to aggregate the raw dataset, we perform two main tasks: (i) identifying relevant search keywords and (ii) conducting automated queries of the dataset across five social media platforms using these targeted search keywords. We provide further details below.

~\textbf{}

\BfPara{Donation Keywords Identification} 
During our incubation phase, we manually reviewed online donation solicitations. We found 14 key terms frequently used in such solicitations, such as \emph{givebetter}, \emph{fund}, \emph{help}, \emph{act of kindness}, \emph{support}, \emph{charity}, \emph{donate}, \emph{donation}, \emph{donor}, \emph{awareness}, \emph{giving}, \emph{foundation}, \emph{contribute}, and \emph{helpsomeone}. These terms were linked with specific causes such as \emph{cancer}, \emph{earthquake}, \emph{firefighters}, \emph{police}, \emph{veterans}, \emph{animals}, \emph{hunger}, \emph{Ukraine}, \emph{Christmas}, and \emph{COVID-19}. Overall, we developed 78 keywords to search relevant posts and profiles across various social media platforms.

~\textbf{}

\BfPara{Data Collection} Utilizing API services\cite{TwitterUserDetailAPI,TwitterTimelinesAPI,InstagramScraper,TelegramApifyScraper,TelegramTelemetrScraper,YouTubeScraper,FacebookScraper}, we gathered data across \emph{X}, \emph{Instagram}, \emph{Facebook}, \emph{Telegram}, and \emph{YouTube} using the formulated keywords. We conducted three separate data searches for each social media platform from 2024-03-03 to 2024-05-15. In total, we collected 151,966 accounts and 3,053,333 posts from five social media platforms. Additionally, we retrieved profile metadata for each account, including name, description, links, profile image, timelines posts, and other publicly available information. A detailed breakdown of the raw data is presented in~\autoref{table:overview_social_media}.

\subsection{Fraudulent Donation Filtration}
After collecting data from 151,966 accounts and 3,053,333 posts across five social media platforms, we conduct data curation. This process involves two primary steps: (i) pre-processing the raw data to confirm it pertains to donation-related contexts, and (ii) filtering candidates associated with donation-based abuses. The following outlines the various steps involved in our data curation techniques to ensure the accuracy of our findings.

\begin{table}[t]
    \small
    \caption{Overview of the raw dataset from five social media platforms. Our dataset reveals that \emph{Telegram} has the highest number of accounts and posts compared to the others.}
    \centering
    \begin{tabular}{lrr}
        \toprule
        \rowcolor{gray!0}
        \multicolumn{1}{c}{\textbf{Social Media}} & \multicolumn{1}{c}{\textbf{Accounts}} & \multicolumn{1}{c}{\textbf{Posts}} \\
        \midrule
        Instagram & 1,604 &  136,082\\
        \rowcolor{gray!10}
        Facebook & 10,607 & 29,349\\
        \rowcolor{gray!0}
        X & 23,871 & 280,789\\
        \rowcolor{gray!10}
        YouTube & 30,482 & 54,314\\
        \rowcolor{gray!0}
        Telegram & 85,402 & 2,552,799\\
        \rowcolor{gray!10}
        \bottomrule
        All & 151,966 & 3,053,333\\
        \bottomrule
    \end{tabular}   
    \label{table:overview_social_media}
\end{table}

\subsubsection{Pre-Processing on Raw Data} In the pre-processing technique we perform filtrations by donation solicitation posts. During our manual analysis of the collected data, we found that API responses often contained irrelevant content. For example, searches using keywords like \emph{donate cancer} yielded results that were not specifically about donations but included general cancer-related content or unrelated donation activities. To address this, we introduced a context check for each account and its associated posts to verify if the content was relevant to donation activities. Using the Large Language Model (GPT-4o)~\cite{openAI}, we developed a prompt injection to identify whether a given post was relevant to the donation context (see~\autoref{sec:prompt_engineering_donation_context}). This filtering process excluded accounts that were unrelated to the donation context: 25.56\% (410/1,604) from \emph{Instagram}, 79.84\% (8,469/10,607) from \emph{Facebook}, 20.77\% (4,959/23,871) from \emph{X}, 80.93\% (24,670/30,482) from \emph{YouTube}, and 89.12\% (76,111/85,402) from \emph{Telegram} were filtered. Across all five social media platforms, this filtering removed 75.42\% (114,619/151,966) of accounts and 82.45\% (2,517,489/3,053,333) posts associated with these accounts from our raw dataset. We then applied security risk engine-based flagged association to the remaining 24.57\% (37,347/151,966) accounts and their 17.54\% (535,844/3,053,333) posts related to the donation context to identify candidate scam accounts.

\subsubsection{Data Filtration and Labelling}
\label{sec:seeds_filtration}
To label an account as a donation solicitation scam, we apply two criteria: (i) the account solicits donations through publicly engaged posts, and (ii) the account's communication channels or profile metadata include elements flagged by security risk engines. If both conditions are met, the account is labeled as a candidate for donation solicitation scam. For example, if a social media profile solicits donations and includes a fraudulent email, phone number, or links to websites flagged by Anti-Phishing Engines as phishing URLs or malicious emails, we categorize it as a donation solicitation fraudster. Further details on the filtering and data labeling techniques are provided below.

\textbf{}

\BfPara{Phishing URLs} We observed that social media profiles often include external websites or URLs in their bio sections. For each profile, we analyze the metadata to check for the presence of any URLs or domains. Using the \emph{VirusTotal} API~\cite{virusTotal}, we evaluate whether these URLs are flagged as phishing or scam sites. To ensure accuracy, we only consider URLs or domains as potential candidates if they are flagged by at least two security risk engines from \emph{VirusTotal}. Accounts or posts containing URLs flagged by \emph{VirusTotal} are marked for further scam donation abuse analysis. 

In total, we identified 118,735 URLs within the profile metadata, and 0.95\% (1,128/118,735) of these distinct URLs were flagged by at least one of the \emph{VirusTotal} security risk engines, spanning 2,345 social media accounts. Of the 1,128 flagged URLs/domains, only 22.34\% (252/1,128) were flagged by two or more security risk engines. A manual review of 5\% of the URLs, both single-flagged and multi-flagged, revealed that single-flagged URLs/domains were often false positives or unknown, while those flagged by two or more engines were found to be reliable. To mitigate potential false positives, we labeled accounts containing 0.21\% (252/118,735) of URLs/domains as candidate accounts linked to 369 social media profiles that were flagged by multiple security risk engines from \emph{VirusTotal}.

\textbf{}

\BfPara{Abusing Email Addresses and Phone Numbers} We observe that social media profiles often include communication methods such as email addresses and phone numbers in their bio-data to facilitate user contact. To assess the reliability of these communication methods, we used third-party API services to check the fraud score of the provided email addresses~\cite{ipQualityEmailAPI} and phone numbers~\cite{ipQualityPhoneAPI}. Social media profiles with communication methods having a fraud score greater than $85\%$ were marked as candidates for further analysis. We set an 85\% threshold based on the providers' high-risk validation, which indicates a strong association with fraud or high-risk activity for the given account. Out of 7,752 email addresses found in our pre-processed data, 2.90\% (225/7,752) distinct email addresses were flagged with high-risk / fraud emails associated with 257 social media accounts. Similarly, out of 9,791 phone numbers found in our pre-processed data, we identified 1.37\% (135/9,791) fraud phone numbers associated with 201 social media accounts. 

\textbf{}

In a nutshell, starting with 151,966 accounts and 3,053,333 posts from five social media platforms, we applied two filtration techniques: (i) Initially removing non-donation-based contexts, and (ii) Further curating the data based on fraud risk engine-flagged URLs/domains, phone numbers, and emails. As a result, our dataset for donation-based scams includes 832 social media profiles. This means we filtered out 99.45\% (151,134/151,966) of the accounts from raw dataset accounts. We acknowledge that our conservative filtering approach may have excluded some donation scam accounts. However, as pioneers in the large-scale study of fraudulent donation scams, our goal was to build a solid foundation using known seed data to reduce potential false positives. Additionally, in ~\autoref{sec:data_set_evaluation_and_discussion}, we explore the data evaluation and the efficacy of scam filtration of our approach. 

\subsection{Tracking and Scam Analysis}

The third component, tracking and scam analysis, focuses on evaluating data from scammers' profile metadata and engagement posts. We analyze profile metadata to investigate the scammers' associations with flagged email addresses, URLs, and phone numbers identified by fraud detection engines. Additionally, we examine engagement posts to understand scammers' interactions and operational methods to show how scammers solicit donations via financial payment methods such as \emph{PayPal}, cryptocurrency addresses, survey forms, and crowdfunding services. By analyzing data from these sources, we provide details on scam operations and the connections between scam accounts across multiple platforms beyond originating social media platforms.

For the rest of the section organization, we provide -- an overview of donation abuse in~\autoref{sec:donation_abuse_overview}; profile content and association in ~\autoref{sec:profile_content_and_association}; fraudulent donation solicitations topologies in~\autoref{sec:post-clustering}; evaluation of scammer's profile picture in~\autoref{sec:profile-clustering}; sentiment analysis of interacted comments in~\autoref{sec:sentiment_analysis}; scammer operations and network analysis in~\autoref{sec:scam_network analysis}; and tracking of scamming payment profiles in~\autoref{sec:financial_tracking}. Additionally, we provide recommendations for mitigating and proactively blocking these fraudulent accounts in \autoref{sec:recommendations}.
\begin{table*}[tb]
    \small
    \caption{Summary of scammers' posts and communication channels. This table shows our findings on donation-based abuses identified by analyzing profile metadata and engagement posts from five social media platforms. For each communication channel—email, phone, and URLs—we perform queries to determine if security risk engines flag the communications.}
    \centering
    \begin{tabular}{lrrrrrr}
        \toprule
        \rowcolor{gray!0}
        \textbf{Platforms} & \textbf{Fraud Email/Accts.} & \textbf{Fraud Phone/Accts.} & \textbf{Malicious URL/Accts.} & \textbf{Distinct Posts} & \textbf{Total Posts} & \textbf{Scammers}\\
       \midrule
        Instagram & 0 & 1/12 & 5/44 & 987 & 4,606 & 56 \\
        \rowcolor{gray!10}
        Facebook & 147/148 & 12/14 & 4/4 & 322 & 323 & 164 \\
        \rowcolor{gray!0}
        Telegram & 8/8 & 63/84 & 14/86 & 6,049 & 6,075 & 188 \\
        \rowcolor{gray!10}
        YouTube & 47/78 & 28/58 & 115/70 & 180 & 200 & 200 \\
        \rowcolor{gray!0}
        X & 23/23 & 32/33 & 114/165 & 6,520 & 6,526 & 224 \\
        \rowcolor{gray!0}
        \bottomrule
        Total (Distinct) & 225/257 & 136/201 & 252/369 & 14,058 & 17,730 & 832 \\
        \bottomrule
    \end{tabular}   
    \label{table:scam_donation_fraud_overview}
\end{table*}

\section{Scam Donation Abuse Overview}
\label{sec:donation_abuse_overview}

In this section, we provide an overview of fraudulent communication channels collected from scammers' profile metadata and engagement posts. In ~\autoref{table:scam_donation_fraud_overview}, we summarize these findings by social media platform. The first column lists the five social media platforms we studied. The second, third, and fourth columns show the number of fraudulent channels associated with the scamming accounts. The fifth and sixth columns provide the distinct and total posts identified in the context of donation scams, and the seventh column shows the overall number of scammers soliciting donations.

\begin{figure}[t]
\centering
\includegraphics[width=.31\textwidth]{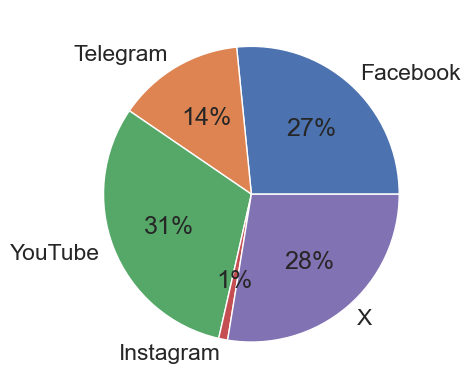} \hfill
\caption{Distribution of security risk engines flagged communication channels (email, phone number, and URLs) across social media platforms. In this pie chart, we show the total number of scamming channels that were flagged by security risk engines identified across five social media platforms, with 31\% of the total communication channels accounting from the \emph{YouTube} platform.}

\label{fig:pie_scamming_channels}
\end{figure}

In total, we identified 225 fraudulent emails, 136 fraudulent phone numbers, and 252 malicious URLs shared by 832 scammers across 17,730 posts and profile metadata. Among these fraudulent communication channels, scammers showed a strong preference for URLs, which accounted for 41.10\% (252/613), often directing victims to external websites for donations. The remaining channels included emails at 36.74\% (225/613) and phone numbers at 22.21\% (136/613). In ~\autoref{fig:pie_scamming_channels}, we illustrate scamming channels by each social media profile, and below, we highlight key findings for each platform studied.

\BfPara{Instagram} In our study, 6.73\% (56/832) of scammers operated on \emph{Instagram}, the lowest count among the platforms analyzed. These scammers preferred using malicious URLs for donation fraud over emails or phone numbers. Among the 56 scamming accounts, we found no fraudulent emails, one fraudulent phone number, and 5 malicious URLs, which appeared in 25.97\% (4,606/17,730) of posts and profile metadata. Notably, these scammers frequently duplicated posts to solicit donations; of the 4,606 posts reviewed, 78.57\% (3,619/4,606) were duplicates.

\BfPara{Facebook} Among the five social media platforms, although \emph{Facebook} had the second-lowest number of scammers at 19.71\% (164/832) and the fewest posts at 1.82\% (323/17,730), it accounted for the highest percentage of fraudulent emails—57.19\% (147/257) of all identified fraud communication channels. This suggests that scammers on \emph{Facebook} were more inclined to engage in donation-based fraud through emails rather than using fraudulent phone numbers or malicious URLs.

\BfPara{Telegram} In our study, \emph{Telegram} had the second-highest post count at 34.26\% (6,075/17,730) and accounted for 22.59\% (188/832) of the scammers. Among the 85 distinct fraudulent communication channels linked to these 188 scamming users, phone calls were the preferred method, making up 74.11\% (63/85). Since \emph{Telegram} is widely used for text messaging and phone calls, scammers on this platform were most likely to connect with victims through phone calls or direct messages.

\BfPara{YouTube} On \emph{YouTube}, 24.03\% (200/832) of scammers were identified, the second-highest after \emph{X}. Among the 190 fraudulent communication channels used by these 200 scammers, external URLs were the most common, accounting for 31.16\% (115/369). Emails followed as the second most used method at 20.88\% (47/225), with phone calls close behind at 20.58\% (28/136).

\BfPara{X} Overall, our study found that the \emph{X} platform is the most favored among scammers, comprising 26.92\% (224/832) of all scammers. Among the 169 fraudulent communication channels identified on \emph{X}, 67.45\% (114/169) were malicious URLs, making them the most common method for donation abuse. Similarly, 36.80\% (6,526/17,730) of the scamming posts featured a significant proportion of malicious URL sharing at 45.23\% (114/252). The findings indicate that fraudulent profiles on \emph{X} prefer using malicious URLs over emails and phone numbers to solicit fake donations.

\textbf{}

\begin{mdframed}[style=insightstyle]
\textbf{Key Takeaways.} 
Through the study of abusive communication channels, we identify that scammers use social media platforms as originating sources, and direct victims to use external channels such as fraud email, phone calls, and URLs to further contact. As URLs provide easy fraud mechanics compared to email and phone calls, scammers prefer URLs as the highest compared to others asking victims to donate via external sites. 
\end{mdframed}

\begin{figure*}[t!]
    \centering
    \subcaptionbox{Scammer Interactions.\label{fig:si_at}}{\includegraphics[width=0.31\textwidth]{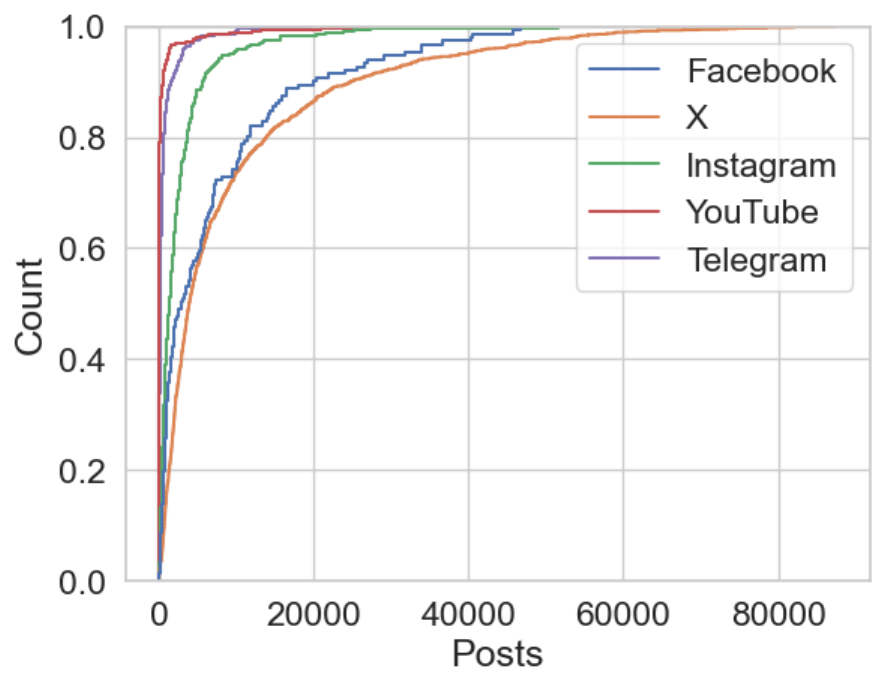}}\hfill
    \subcaptionbox{Following Count.\label{fig:si_ap}}{\includegraphics[width=0.30\textwidth]{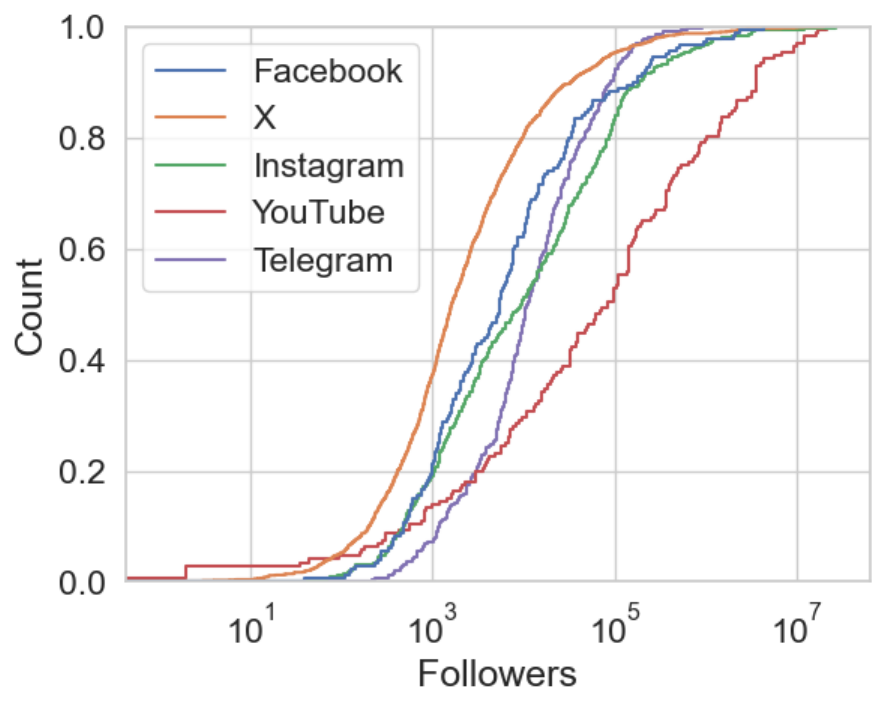}}\hfill
    \subcaptionbox{Profile Creation Date.\label{fig:si_ac}}{\includegraphics[width=0.35\textwidth]{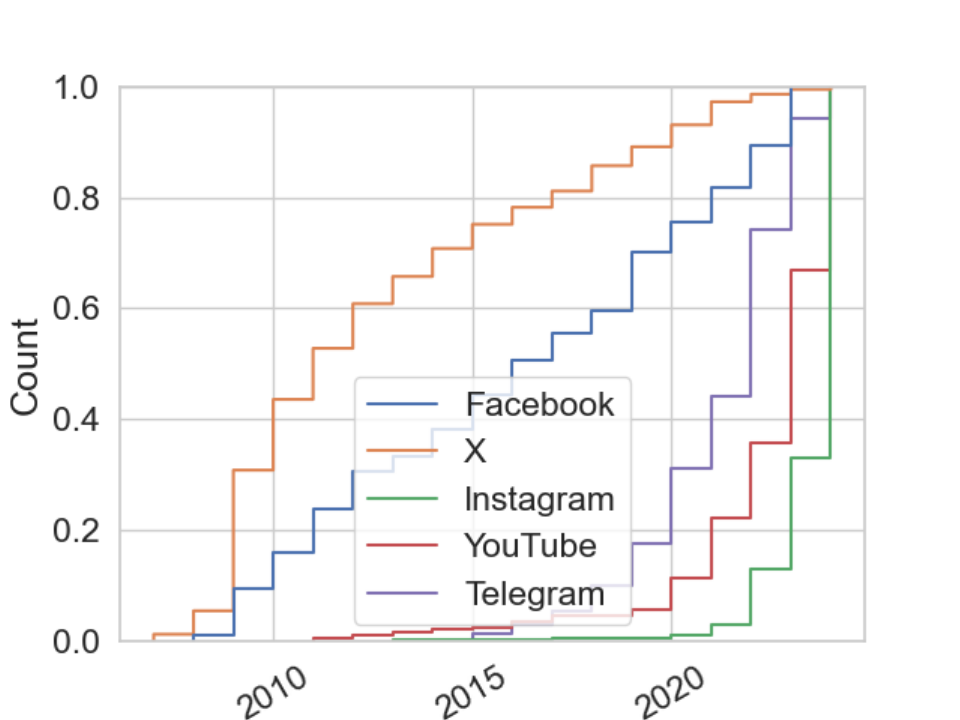}}

    \caption{CDF Engagement and age of scammer profile from each of the social media platforms -- \autoref{fig:si_at} shows the engagement of scammer via posts, \autoref{fig:si_ap} shows the {\em following} count of scammers and \autoref{fig:si_ac} shows the age of scammers based on profile creation date from each of the social media platforms that we studied.}
    \label{fig:scammer_profile_evaluation}
\end{figure*}
\section{Profile Content and Association}
\label{sec:profile_content_and_association}

In this section, we dive deep into scammers' techniques to create profiles that attract potential victims on social media platforms. We conduct a thorough analysis of six key aspects: post engagement, follower count, account age, location settings, categorical representation, and account monetization. In \autoref{fig:scammer_profile_evaluation}, we present a CDF graph showing scammers' engagement through posts, follower count, and account creation dates, and below we provide further details on profile content and associations.

\BfPara{Description/Bio} Scammers engaged in fraudulent donations were found to use various tactics in their profile descriptions. These descriptions provide a brief message to visitors. We found that 95.31\% (793/832) of scammers contained profile descriptions that included messages related to emotional manipulation, credibility, authentication, details about the donation campaign, or appeals to generosity. The remaining 4.68\% (39/832) of scammers were found to lack any description or bio information.

\BfPara{Posts Engagement} Out of 17,730 posts collected from five social media platforms, the overall median post interaction across all platforms was 709. The median post interactions for each platform were: \emph{X} (4,775), \emph{Instagram} (1,307), \emph{YouTube} (2), \emph{Telegram} (147), and \emph{Facebook} (2,960). Our results indicate that scammers are more likely to engage on \emph{X} compared to other platforms, whereas \emph{YouTube} is less favored for engagement through posts. 

\BfPara{Followers Engagement} The median follower count across the five social media platforms was 3,345. For each platform, the median follower counts were - \emph{X} (1,621), \emph{Instagram} (9,361), \emph{YouTube} (82,500), \emph{Telegram} (10,809), and \emph{Facebook} (5,449). Our result indicates that users are more inclined to follow scammers in video-based donation contexts compared to post-based ones. Since videos are generally more engaging than posts, scammers may find it easier to attract and retain followers through video content.

\BfPara{Account Age}  Our analysis of fraudulent social media profile ages reveals that scammers often use either harvested or aged profiles. The median creation date for all social media profiles was 2018. Specifically, the median ages for each platform were: \emph{X} (2011), \emph{Instagram} (2024), \emph{YouTube} (2023), \emph{Telegram} (2022), and \emph{Facebook} (2016). This indicates that scammers are more likely to utilize older accounts on \emph{X} while utilizing newer accounts on Instagram. 

\BfPara{Location} We identified 50.12\% (417/832) of scammers with 210 distinct geo-location sets as part of their profile information. The top three countries represented were \emph{Russia} (62), the \emph{USA} (37), and \emph{India} (29). It is important to note that geo-location is an optional field and does not necessarily reflect the scammers' actual locations, as it is often populated with unrelated names. For example, the location name ~\emph{global}, although not a real location, had the highest count, with a total of 77.

\BfPara{Categorical Representation} We identified that 49.87\% (415/832) of scamming accounts featured 115 distinct categories or affiliations in their profiles. Among these, the top three categories included: \emph{Non-Profit Organization} (62), \emph{Charity Organization} (22), and \emph{Non-Governmental Organization (NGO)} 17. The remaining 50.12\% (417/832) were found to display missing categorical representation. 

\BfPara{Account Monetization} We found that 39.66\% (330/832) of scamming accounts across four social media platforms (\emph{Facebook} (22/164), \emph{Instagram} (30/56), \emph{X} (218/224), and \emph{YouTube}) (60/200), opted for business or advertisement features. This enables these accounts to monetize their presence and allows the platforms to display advertisements. Accounts with higher engagement levels generally gain more from opting into these business features. Notably, 97.32\% of scamming accounts on the \emph{X} platform had the highest participation in business or advertisement features.

\textbf{}

\begin{mdframed}[style=insightstyle]
\textbf{Key Takeaways.} 
Scammers were found to use older social media accounts to launch donation abuse campaigns. We suspect these are rather harvested accounts. Scammer's geo-location data shows diverse representations of top countries including Russia and the USA, though these locations are often misleadingly named. Moreover, scammers often masquerade under popular affiliations and opt-in for business/advertisement features, allowing for monetization through advertisements.
\end{mdframed}
\section{Fraud Topologies: Anatomy of Fraudulent Donation Solicitations}
\label{sec:post-clustering}
In this section, we provide the categories of fraudulent donation solicitations that scammers perform through posts.  We provide a technical overview and the findings of the scam clusters below.

\BfPara{Technical Overview} We clustered donation solicitation posts from 832 scamming profiles excluding non-English content. In total, we analyzed 17,706 posts across five platforms: \emph{X} (6,526), \emph{Instagram} (4,583), \emph{Telegram} (6,075), \emph{Facebook} (322), and \emph{YouTube} (200). For language identification and filtering, we use the CLD2 library~\cite{cld2-cffi}. We then vectorized the posts using the \emph{all-mpnet-base-v2} sentence transformer model~\cite{Reimers2019SentenceBERTSE}. Subsequently, we processed the posts using the BERTopic library~\cite{grootendorst2022bertopic} to remove redundant information, such as stop words. We combined UMAP~\cite{McInnes2018} and HDBSCAN~\cite{McInnes17hdbscan} for clustering, followed by the KeyBERT~\cite{grootendorst2020keybert} model to refine topic representations within each cluster. 
% Additional details on text processing and clustering are available in the supplementary material~\autoref{appendix:post-clustering}.

In the hyperparameterization process for UMAP, default values from the BERTopic library~\cite{grootendorst2022bertopic} were employed. 
Specifically, we configured UMAP with \texttt{n\_neighbors=15}, \texttt{min\_dist=0.0}, \texttt{n\_components=5}, and cosine similarity. 
We then set the \texttt{random\_state} variable to a fixed value of $42$ to preserve the reproducibility of our code.

For HDBSCAN, we chose \texttt{min\_cluster\_size=10} and used the Euclidean metric for clustering. To refine the clustering outcome, we adjusted \texttt{min\_samples=50} to reduce the resulting number of clusters. 
Additionally, the default BERTopic method for outlier reduction (\texttt{reduce\_outliers}) was applied to minimize the presence of outlier samples in the clustering results.
Finally, we employed a standard evaluation metric, i.e., silhouette score~\cite{Shahapure2020ClusterQA}, and visual inspection of resulting clusters to assess the quality and validity of the clustering outcomes.

\textbf{Clustering Results.}  We conducted a manual qualitative analysis of prominent scam categories identified in our findings. 
Out of the 62 clusters identified through our clustering pipeline, we present below an analysis of the top 10 clusters based on engagement through posts where scammers solicit fraudulent donations.

\begin{itemize}

\item \BfPara{Urgent Support} We observe that scammers frequently target specific donation days or weeks to create a sense of urgency, often setting rapidly approaching deadlines. A common tactic involves urging social media users to complete survey forms or to visit an external website before the donation period ends. We identified 185 scammers asking for urgent support fraudulent donations through $951$ posts, which comprised the highest numbers of scammers and post-interactions in our study.

\item \BfPara{Animal Rescue} In the context of animal rescue abuse, scammers target individuals by posing as representatives of legitimate animal rescue organizations to establish credibility. These fraudulent posts solicit donations under the guise of supporting animal welfare causes, asking for contributions to help save and care for animals in need. In this cluster, we identified 125 scammers asking for fraudulent animal rescue donations via 679 posts.

\item \BfPara{Disaster Relief} We observe scammers often exploit the impact of disaster relief to solicit fraudulent donations. In this category, scammers act as legitimate organizations or affiliations preying on those looking to support natural disaster victims. We identified 87 scammers asking for disaster relief fraudulent donations through 426 posts. 

\item \BfPara{Event and Activities Support} Scammers in this category exploit popular events to solicit fraudulent donations, leveraging the excitement and urgency to support the occasions. The scammer was often found to craft persuasive messages appealing to participants' emotions and sense of community, urging them to contribute financially. In this cluster, we identified 80 scammers asking for fraudulent donations via 427 posts.

\item \BfPara{Crypto Scams} In the context of crypto donation abuse, we identify that scammers exploit the growing popularity and perceived anonymity of cryptocurrency to solicit fraudulent donations. They take advantage of the novelty and complexity of cryptocurrency, making it appealing for users to either participate in charity-related philanthropic support or take part in free crypto token giveaways. In this cluster, we identified 57 scammers asking for fraudulent crypto donations via 111 posts.

\item \BfPara{Holiday/Seasonal Spirit} Scammers in this category exploit the holiday or seasonal spirit of generosity to make fraudulent donation requests. These scams are often focused on children and families in need. In this cluster, we identified 54 scammers soliciting fraudulent donations through 382 posts.

\item \BfPara{Education/Research Support} In this category, we observe scammers exploit the education sector by targeting individuals with fraudulent donation requests related to scholarships, educational research, and student support. These scams often pose as associations to institutions or charitable initiatives, appealing to the goodwill of alumni, faculty, and the general public. We identified 49 scammers soliciting fraudulent donations through 92 posts in this category. 

\item \BfPara{Ticketing and Offer Exchange} We observe that scammers in this category claim to need tickets or offer ticket exchanges as part of fraudulent ticket donations. Scammers perform potential disguises as potential donors to fraudulent websites or request personal information under the guise of facilitating a ticket donation. By creating a sense of urgency and community solidarity, they deceive well-meaning fans into buying a sold-out ticket or providing financial support for a particular event through ticket purchases. In this cluster, we identified 46 scammers asking for fraudulent ticket donations via 81 posts.

\item \BfPara{Narcissistic Abuse Support} In this category of fraudulent donation solicitations, scammers target individuals by asking for support for abused groups, particularly those affected by narcissistic abuse. Their tactics include raising awareness and soliciting donations for victims of war, domestic violence, and psychological abuse. In this cluster, we identified 40 scammers asking for fraudulent donations via 63 posts.

\item \BfPara{Medical} In medical-related fraudulent donation requests, scammers are found to solicit funds for various medical causes, such as covering the medical expenses of a critically ill patient, supporting medical research, or providing medical care for disadvantaged groups. They often impersonate medical institutions to add legitimacy to their appeals. In this cluster, we identified 36 scammers asking for fraudulent medical-related donations via 265 posts. 
\end{itemize}

\textbf{}

\begin{mdframed}[style=insightstyle]
\textbf{Key Takeaways.} 
Our analysis of post-clustering uncovered several scam categories of fraudulent donations performed by social media profiles. These include urgent appeals with specific deadlines, schemes tied to events, holiday-themed solicitations for families and children, and deceptive campaigns masquerading as education and research support. Furthermore, scammers exploit disaster relief efforts, victims of abuse, animal rescue, and medical issues, presenting themselves as legitimate fundraisers while seeking fraudulent donations.
\end{mdframed}

\section{Evaluation of Scammer Profile Picture}
\label{sec:profile-clustering}
In this section, we provide an evaluation of the scammer's choice of profile picture while soliciting donations across multiple social media platforms. We provide a technical overview and the findings of the profile picture evaluation below. 

\BfPara{Technical Overview} Using unsupervised clustering to identify patterns and relationships among these images, we examine the profile pictures of scammer accounts. 
Following the methodology outlined in \cite{acharya2024conning}, we collected profile pictures and employed the pre-trained visual model CLIP~\cite{Radford21CLIP} for feature extraction.
For each profile picture, we extracted the CLIP token embeddings and rescaled the images to a resolution of $224\times224$ pixels to match the input size used during the model's training~\cite{Radford21CLIP}. These embeddings were then visualized using Uniform Manifold Approximation and Projection (UMAP)~\cite{McInnes2018}.
To identify clusters and eliminate anomalies, we applied standard clustering algorithms: HDBSCAN~\cite{McInnes17hdbscan} and single-linkage hierarchical clustering~\cite{Hastie01Hierarchical}. 
Below we provide additional detailed information on the chosen hyperparameters and clustering validation, and the results of our findings.

\BfPara{Clustering Hyperparameters Selection}
During hyperparameters selection, we employ standard evaluation metrics, i.e., silhouette score~\cite{Shahapure2020ClusterQA} and Calinski-Harabasz score~\cite{maulik2002performance}, and visual inspection of resulting clusters to assess the quality and validity of the clustering outcomes. 
For both UMAP and DBSCAN, we systematically tuned their hyperparameters to optimize clustering pipeline performance and obtain meaningful and reliable results. 
To this end, we considered a wide range of hyperparameter configurations. 
Specifically, for UMAP, we let the \texttt{n\_neighbors} hyperparameter vary in the intervals $[3, 100]$ and set the \texttt{n\_components} equals to $2$ to visualize the clusters. 
Regarding DBSCAN, we let the \texttt{min\_cluster\_size} and \texttt{min\_dist} vary in the intervals $[5, 100]$  and $[1e-02, 1]$ respectively. 
The resulting investigation involved $2,500$ configurations of these hyperparameters, identifying the configuration \texttt{n\_neighbors}=$15$, \texttt{n\_components}=$2$, \texttt{min\_dist}=$0.1$, and \texttt{min\_cluster\_size}=$20$ as the most reliable, according to their silhouette and Calinski-Harabasz scores, for our clustering pipeline.

\begin{table}[tb]
\caption{Clustering analysis of scammers' profile pictures and their distribution across the five social media platforms. Our result reveals that scammers in the {Association-Logos} category were found to be the highest and utilize association logos to solicit donations.}
\label{tab:profile-picture-scammers}
\resizebox{\columnwidth}{!}{%
\setlength\tabcolsep{1.5pt}
\begin{tabular}{@{}lllllll@{}}
\toprule
\textbf{\small Cluster Label } & \textbf{Count} & \textbf{Facebook} & \textbf{ X} & \textbf{Telegram} & \textbf{Instagram} & \textbf{Youtube} \\ \midrule
Associations Logos & 240 (29.13\%)  & 79& 102 & 34& 24& 1\\
\rowcolor{gray!10} Male/Female& 133  (16.14\%)  & 14& 48 & 42& 12 & 17\\
Video Clips& 110  (13.35\%)  & 0& 0 & 3& 1& 104\\
\rowcolor{gray!10} Games \& Cartoon& 103  (12.50\%)  & 12& 33 & 49& 1 & 7\\
Politics/War& 97(11.77\%)& 0 & 0& 33 & 1  & 63\\
\rowcolor{gray!10} Pets & 85(10.31\%)& 48 & 17& 0 & 16 & 4\\
Low-Resolution& 37  (4.49\%)& 11 & 16& 9 & 1  & 0\\
\rowcolor{gray!10} Crypto Coins& 19(2.31\%)& 0 & 3& 16 & 0 & 0\\
 \midrule
\textbf{Total}& 824& 164& 219& 186& 56& 196 \\ \bottomrule
\end{tabular}
}
\end{table}

\BfPara{Clustering Results}
We present the results of our clustering analysis on $824$\footnote{We excluded $8$ images due to unsupported formats (e.g., non-JPEG or non-PNG)} scammer profile images in~\autoref{tab:profile-picture-scammers}. From the analyzed dataset, we identified seven common categories of profile pictures used by scammers: \emph{Association Logos}, \emph{Male/Female}, \emph{Video Clips}, \emph{Games \& Cartoon}, \emph{Politics/War}, \emph{Pets}, \emph{Low-Resolutions}, and \emph{Crypto Coins}. 
Our results show that $29\%$ of scammers use \emph{Association Logos} as their profile pictures, often featuring logos from various groups such as pacifist organizations, religious institutions, private companies, or even the \emph{Ukraine} flag. About $16\%$ of scammers use \emph{Male} or \emph{Female} profile pictures, while $13\%$ fall into the \emph{Video Clips} category, using video snapshots as their profile images. The \emph{Games \& Cartoon} category, comprising $12\%$ of scammers, includes images of video game characters, anime protagonists, and memes.
Additionally, $11\%$ of scammers employ \emph{Political War} images, such as screenshots from political news, military actions, or propaganda. The \emph{Pets} category ($10\%$) features images of animals, mostly cats and dogs, as well as pet-related activities. Scammers using low-quality images belong to the \emph{Low-Resolution} cluster ($4\%$), where the content is difficult to discern. Finally, $2\%$ of scammers fall into the \emph{Crypto Coins} cluster, which includes images of cryptocurrencies, and wallet logos.

Our analysis of scammers' profile images revealed that they often aim to emotionally manipulate users by featuring images of pets, war, educational organizations, or religious themes. Additionally, in Appendix, \autoref{fig:scammers_cluster_1}-\ref{fig:scammers_cluster_3}, we show a subset of $50$ scammer profile pictures from \textit{Association-Logos}, \textit{Male}, \textit{Female}, \textit{Games-Cartoon}, \textit{Politics-War}, \textit{Pet Associations}, and \textit{Pets} clusters. 
Notably, the content within the clusters we identified is cohesive and coherent with our assigned label. Complementary, in \autoref{fig:scammer_cluster_miscellaneous}, we illustrate samples coming from the \textit{Miscellaneous} cluster, which contains a mixture of pictures that have been considered anomalous by our clustering algorithms.

\textbf{}

\begin{mdframed}[style=insightstyle]
\textbf{Key Takeaways.} 
 Through profile image analysis, we identify patterns and tactics used by scammers to create a deceptive online presence. Our analysis revealed that scammers predominantly use association logos, male and female images, political war, and game/cartoon characters to appear credible. Such insights are valuable for developing targeted measures to detect and counteract fraudulent activities, improving online security across social media platforms.
\end{mdframed}
\section{Sentiment Analysis of Public Comments}
\label{sec:sentiment_analysis}
In this section, we conduct sentiment analysis between users and scammers. We focused specifically on YouTube due to its unique video-based interaction format. Users often engage with videos as directed by the content, which differs from textual posts found on posts-based interacting platforms (~\emph{X}, ~\emph{Instagram}, ~\emph{Facebook}, and ~\emph{Telegram}). We collected 3,676 distinct comments from 364 scamming YouTube channels. 

\BfPara{Technical Overview} For sentiment analysis, we utilized the Llama3-8B model based on its popularity as the start of an art open-source model on benchmark sentiment analysis. Our comment categorization was based on predefined sentiments: \textit{Gratitude}, \textit{Action}, \textit{Anger}, \textit{Abuse}, and \textit{Neutral}. We provide the prompt detail to these five sentiments in ~\autoref{fig:llama-prompt}.

\BfPara{Sentiments Results} We provide detailed results of sentiment analysis of post engagement between users and scammers during the lifecycle of fraudulent donation solicitations as below.

\BfPara{Gratitude}
In the \emph{Gratitude} category, we measured comments expressing gratitude, relief, or thankfulness. We found that 53.73\% of the comments reflected gratitude. We observe that scammers frequently try to thank those who have already donated and solicit others to make additional fraudulent donations. An example of a scammer's gratitude is shown below. 

\textbf{}

\noindent \emph{Thank you, every single donation matters, even if you can't donate.}

\textbf{}

\noindent \emph{God bless everybody involved in the rescue and care of this beautiful dog family.} 

\textbf{}

\noindent \emph{7 hours and already \$50,000 donated... Thank you for improving the lives of so many others.} 

\textbf{}

\begin{figure}[t]
    \centering

\begin{tcolorbox}[
    colframe=black!70,    % Border color
    colback=black!5,      % Background color
    coltitle=white,      % Title color
    fonttitle=\bfseries,
    title=Classifier Task Description,
    sharp corners,
    boxrule=1pt
]
\textbf{You are a classifier.} Given a \textbf{Comment}, classify it into one of the following categories:

\vspace{0.5em}
\textbf{\textcolor{green!70!black}{Gratitude}}: A comment expressing gratitude, relief, or similar emotions.

\textbf{\textcolor{orange!90!black}{Action}}: A comment that includes awareness, a report, an urgent action, or similar prompts.

\textbf{\textcolor{red!70!black}{Abuse}}: A comment indicating that scammers are engaging in hateful, abusive, fearful, or concerning activities.

\textbf{\textcolor{purple!70!black}{Anger}}: A comment showing that the user is frustrated or angry because they believe YouTube is not taking serious steps to block scam accounts.

\vspace{0.5em}
\textbf{Provide your classification in the following format:}
\begin{itemize}
    \item \textbf{Category:} \textit{"..."}
    \item \textbf{Explanation:} \textit{"..."}
\end{itemize}

\vspace{1em}
\textbf{Examples:}

\begin{tcolorbox}[colback=gray!10, colframe=black!50, sharp corners, boxrule=0.5mm]
\textbf{Comment}: "Thank you so much for addressing this issue! I was really worried."\\
\textbf{Category}: "Gratitude"\\
\textbf{Explanation}: "The comment expresses gratitude and relief for addressing the issue."
\end{tcolorbox}

\begin{tcolorbox}[colback=gray!10, colframe=black!50, sharp corners, boxrule=0.5mm]
\textbf{Comment}: "Everyone needs to report these scammers immediately!"\\
\textbf{Category}: "Action"\\
\textbf{Explanation}: "The comment is a call to action, urging others to report scammers."
\end{tcolorbox}

\end{tcolorbox}
    \caption{\textbf{System prompt for Llama-3.} We instruct-tune Llama-3-8B to classify sentiment in Youtube users comments with a system prompt describing the task and two examples.}
    \label{fig:llama-prompt}
\end{figure}

\BfPara{Action} 
In the \emph{Action} category, we measured comments that include awareness, report, urgent action, or time-based responses. We found that 17.79\% of the comments interacted with scamming videos displayed action. We provide examples of action below.

\textbf{}

\noindent \emph{Donate please, another 7.4 earthquake struck Nepal just now.} 

\noindent \emph{Quality of life and hospice support is imperative. Now that you learned how to make a donation button in PLS DONATE}. 

\textbf{}

\noindent \emph{Most large charities are scams with a fraction of donated money ever reaching those it was gifted for, give to local charities that actually do good work}.

\textbf{}

\BfPara{Anger} In \emph{Anger} category, we measured comment that shows that the user is frustrated or angry because YouTube does not take serious steps in blocking the scamming accounts or scammers. We found that 16.43\% of the dataset typically showed frustration with YouTube’s handling of scam accounts. 

\textbf{}

\noindent \emph{They need to be closed down and thrown in jail for fraud.}

\textbf{}

\noindent \emph{The scam part angers me.}

\textbf{}

\noindent \emph{Contact us about paying them for their scam a** service.}

\textbf{}

\BfPara{Hate} In \emph{Hate} category, we measure engagement in hateful or abusive behavior on interaction. We identified 11.62\% of the highlighted engagement comprised of hateful or abusive behavior. Examples of such hateful comments from scammers are shown below. 

\textbf{}

\noindent \emph{You hate charity because you\'re a cringe Socialist.} 

\textbf{}

\noindent \emph{You should get out there on the streets and do the fuc**ng work.}

\textbf{}

\noindent \emph{You’re a lying imposter you deserve what misfortune that comes your way}.

\textbf{}

\BfPara{Neutral} In \emph{Neutral}, we measure interaction that is not necessarily related to donation-based context or posts. We suspect these neutral comments are rather scripted to gain followers. We found the neutral context as the lowest category comprising 0.43\% of our overall dataset. An example of a neutral comment are shown below.

\textbf{}

\noindent \emph{Fun fact: snakes actually use their tongues to catch scents!}

\textbf{}

\noindent \emph{Line from Seinfeld: \"George likes his Kung Pao SPICY\"}. 

\textbf{}

\noindent \emph{If you are impressed with this video, please support us on Patreon - https://www.patreon.com/Le**cs. It will be a great help for us.}

\textbf{}

\begin{mdframed}[style=insightstyle]
\textbf{Key Takeaways.} 
Our analysis of scammer and user interaction sentiments revealed several key insights. In the \emph{Action} category, comments reflected urgent responses or awareness, with some users advising against taking action due to mistrust of large charities. The \emph{Anger} category showed that comments expressed frustration with YouTube's failure to block scam accounts. In the \emph{Hate} category, interactions involved hateful or abusive behavior, both from scammers and users. Lastly, the \emph{Neutral} category included unrelated, scripted comments and motives to gain followers. This indicates that comment-based interactions are lucrative channels of operations for scammers, offering interactive video-based solicitations for donations.
\end{mdframed}

\section{Scammer Network Analysis}
\label{sec:scam_network analysis}
In this section, we explore how scammers operate across multiple social media platforms, focusing scam cycle and modus operandi. We detail the fraud lifecycle, illustrating how scammers redirect users from one platform to another through tactics such as crowdsourcing, and external links, and share scam channels across multiple profiles. We provide further details below.

\subsection{Operation Beyond Originating Platform}
\label{sec:operation_beyond_origin}
In this section, we specifically focus our analysis on scammers operating beyond the originating platforms and interlinking accounts among multiple platforms. Our analysis primarily covers (i) external platforms that scammers link to their profiles, (ii) donation solicitations via crowdfunding services, and (iii) survey forms. Below, we provide detailed information on each category.

\BfPara{External Communication Channels} Through profile metadata analysis, we found that scammers frequently include details of external platforms in their bio descriptions, linking them to the originating social media platform. We identified two types of external bio links on scamming profiles. The first type links to external websites such as Linktree URLs, which aggregate multiple platforms and related links to the scammer's account. For example, a bio profile linking to \emph{www.linktree.com/scam\_account\_1} was often found to contain various social media accounts associated with scam accounts, such as \emph{www.facebook.com/scam\_account\_f}, and \emph{www.twitter.com/scam\_account\_t}. The main purpose of these accounts is to provide visitors with a choice of platforms for contact. The second type involves direct links to a preferred platform, such as an ~\emph{X} profile containing links to ~\emph{Instagram} or ~\emph{Telegram} as part of the external contact details. 

We observed that 37.5\% (312/832) of scamming accounts included external links in their profiles, with 127 of these accounts linking multiple bio profiles (ex. \emph{Linktree}) to external websites. For accounts with multiple external bio links, we automated the Selenium Python script to gather the associated platforms interlinked with the originating account. In ~\autoref{table:overview_exteral_social_media_platforms}, we present data showing 832 scamming accounts interlinked with 11 different platforms across both categories. Overall, we identified 1,001 distinct external platform accounts linked beyond the study accounts. Among these platforms, the top five most commonly interlinked accounts were related to \emph{YouTube} (48.15\%), \emph {Instagram} (16.58\%), \emph{Facebook} (12.18\%), \emph{Twitter} (8.29\%), and \emph{Amazon} (4.39\%). To gain further insights, we conducted a manual analysis by randomly selecting 100 accounts and visiting each link through a browser. We identified four distinct scam operation techniques: (i) platforms such as YouTube were used for video-based donation requests, (ii) messaging platforms such as \emph{Signal}, \emph{Telegram}, and \emph{WhatsApp} were used for direct communication, (iii) social media platforms like \emph{Twitter}, \emph{Instagram}, and \emph{Facebook} were primarily utilized for post engagement, and (iv) consumer-oriented platforms such as \emph{Amazon} and \emph{Etsy} were exploited by scammers to solicit support through purchases from wishlists or gifts. Thus, starting with 832 scamming accounts from five social media platforms, this technique yielded an additional 1,001 external accounts linked to 11 platforms (9 social media platforms and 2 online e-commerce platforms).

\begin{table}[t]
    \small
    \caption{Overview of the external platforms linked to the scam accounts. In this table, we show scammers interlinking various platforms as part of a scam operation.}
    \centering
    \begin{tabular}{lrr}
        \toprule
        \rowcolor{gray!0}
        \multicolumn{1}{c}{\textbf{Social Media}} & \multicolumn{1}{c}{\textbf{External Linked Accounts}} \\
        \midrule
        YouTube & 482\\
        \rowcolor{gray!10}
        Instagram & 166\\
        \rowcolor{gray!0}
        Facebook & 122\\
        \rowcolor{gray!10}
        Twitter & 83\\
        \rowcolor{gray!0}
        Amazon & 44\\
        \rowcolor{gray!10}
        LinkedIn & 36\\
        \rowcolor{gray!0}
        TikTok & 30\\
        \rowcolor{gray!10}
        Telegram & 25\\
        \rowcolor{gray!0}
        Etsy & 9\\
        \rowcolor{gray!10}
        Signal & 2\\
        \rowcolor{gray!0}
        WhatsApp & 2 \\
        \rowcolor{gray!10}
        \bottomrule
        All (Distinct) & 1,001\\
        \bottomrule
    \end{tabular}   
    \label{table:overview_exteral_social_media_platforms}
\end{table}

\BfPara{Crowdfunding Services} We found that scammers exploit crowdfunding services for donation solicitations. We analyzed the presence of popular crowdfunding service URLs in posts engaged by scammers. As shown in~\autoref{table:overview_crowdfunding_services}, we identified 9.97\% (83/832) of scammers soliciting donations via 77 URLs from seven different crowdfunding services. The top three platforms used were \emph{Patreon} (51.94\%), \emph{Donorbox} (24.67\%), and \emph{Kickstarter} (10.38\%). We conducted a manual review of these 77 URLs by visiting each link in a browser. Out of 77 distinct URLs, 9 links were either inactive or deleted. Among the active URLs, 23/68 had already closed their fundraising campaigns, with amounts raised ranging from \$25 to \$58,180. The remaining 45/68 crowdfunding URLs were found to be actively collecting donations, using three main solicitation methods: (i) minimal payments to join a group as a form of support for the cause (e.g., \emph{Patreon} memberships starting at \$1.70 per month plus tax), (ii) recurring donations such as monthly or annual contributions (\$5, \$25, or higher), and (iii) one-time payments for support (ranging from \$5 to several hundred dollars). Our analysis from the last week of September 2024 identified 3,696 contributors who donated over \$252,620 through 37 active fundraising links. This amount does not include contributors who may have made or are still making donations via membership subscriptions. We suspect scammers are repeatedly defrauding victims through ongoing solicitations observed in our dataset.

\begin{table}[t]
    \small
    \caption{Overview of the crowdfunding services. Our results show that scammers often redirect users from the original social media platforms to seven crowdfunding services.}
    \centering
    \begin{tabular}{lrr}
        \toprule
        \rowcolor{gray!0}
        \multicolumn{1}{l}{\textbf{Crowdfunding Services}} & \multicolumn{1}{c}{\textbf{Scam Accounts}} & \multicolumn{1}{c}{\textbf{Fund Links}} \\
        \midrule
        Patreon & 45 &  40\\
        \rowcolor{gray!10}
        Givebutter & 24 & 6\\
        \rowcolor{gray!0}
        Donorbox & 6 & 19\\
        \rowcolor{gray!10}
        Kickstarter & 4 & 8\\
        \rowcolor{gray!0}
        Indiegogo  & 2 & 2\\
        \rowcolor{gray!10}
        Fundrazer & 1 & 1\\
        \rowcolor{gray!0}
        Rallyup & 1 &  1\\
        \rowcolor{gray!10}
        All (Distinct) & 83 & 77\\
        \bottomrule
    \end{tabular}   
    \label{table:overview_crowdfunding_services}
\end{table}

\subsection{Campaign Detection}
We analyzed shared communication channels, specifically URLs, emails, and phone numbers, among scam accounts to determine whether these channels interlink scam accounts as part of their communication with potential victims. To do this, we aggregated data from abuse candidate scam accounts across all five social media platforms. If a minimum of two scam accounts share a single communication channel, we refer to the given group as a scam campaign shared by the scam accounts. 
\begin{table}[t]
    \small
    \caption{Overview of scammers sharing the communication channels. The table provides a breakdown of clusters and scam accounts from all five social media platforms by individual communication channels.}
    \centering
    \resizebox{\columnwidth}{!}{%
    \begin{tabular}{lrrrrrr}
        \toprule
        \rowcolor{gray!0}
        \multicolumn{1}{c}{\textbf{Channels}} & \multicolumn{1}{c}{\textbf{Min}} & \multicolumn{1}{c}{\textbf{Median}} & \multicolumn{1}{c}{\textbf{Max}} & \multicolumn{1}{c}{\textbf{Cluster}} & \multicolumn{1}{c}{\textbf{Accts.}} & \multicolumn{1}{c}{\textbf{Accts.\%}} \\
        \midrule
        Email & 2 &  3 & 8 & 12 & 44 & 17.12\\
        \rowcolor{gray!10}
        Phone & 2 & 3 & 15 & 21 & 88 & 43.78\\
        \rowcolor{gray!0}
        URL & 2 & 2 & 42 & 41 & 231 & 62.60\\
        \rowcolor{gray!10}
         All (Distinct) & 2 & 3 & 42 & 74 & 355 & 42.66\\
        \bottomrule
    \end{tabular}   
    \label{table:individual_channel_cluster}
    }
\end{table}

We grouped the scam accounts based on individual types of communication channels, such as emails, URLs, or phone numbers. In~\autoref{table:individual_channel_cluster}, we summarize the scam clusters, including the minimum, maximum, and median counts of scam accounts per cluster. Overall, 42.66\% of scam accounts were found to be part of scam campaigns. Among these, URL clusters were the most prevalent, with 41 distinct clusters comprising 231 accounts, while email clusters were the least common, with 12 distinct clusters involving 44 scam accounts. The largest cluster contained 42 scam accounts linked through URLs, while the smallest and median cluster sizes across the three communication types were 2 and 3 accounts, respectively.

\textbf{}

\begin{mdframed}[style=insightstyle]
\textbf{Key Takeaways.} 
Scammers leverage multiple platforms and interlink accounts to broaden their operations, frequently redirecting users through strategic bio links and aggregating various platforms. Our analysis reveals that platforms such as \emph{YouTube}, \emph{Instagram}, and \emph{Amazon} are often exploited for donation requests. Scammers also use crowdfunding services to solicit both recurring and one-time contributions. Moreover, scammers operate in organized clusters, connecting campaigns through URLs, emails, or phone numbers, showcasing their advanced and coordinated methods for targeting victims.
\end{mdframed}
\section{Financial Validation and Tracking Payments}
\label{sec:financial_tracking}
From the profile metadata and post engagements of scammers on five social media platforms, we observed that fraudsters soliciting donations often involve requesting payments via various methods such as ~\emph{PayPal} and cryptocurrency addresses. To further validate these scams' impact, we partnered with ~\emph{PayPal}, and ~\emph{Chainabuse}, sharing 1,898 email addresses with ~\emph{PayPal}, and 142 cryptocurrency addresses with ~\emph{Chainabuse}. Below, we present the findings related to these scamming payment profiles based on feedback from our industry partners.

\BfPara{PayPal's Scam Validation}
From the 1898 email addresses that were shared, ~\emph{PayPal} was able to identify and associate 79.71\% (1513/1898) of these to ~\emph{PayPal} accounts on the platform. Among these identified accounts, 26\% were restricted at some point during their activity on the platform. Within these 26\% restricted accounts, above 50\% had more than one restriction placed throughout their time on ~\emph{PayPal}, and 42\% were currently restricted at the time of data sharing. Finally, based on the overall restrictions placed on these accounts, the top reasons were (i) KYC (Know your Customer) \& Compliance concerns, and (ii) Risky Operations like Unauthorized Account Access or Creation.

\BfPara{Chainabuse Scam Validation} 
Out of 142 addresses, 21.83\% (31/142) were identified as invalid. We are unclear as to why scammers provide invalid cryptocurrency addresses when soliciting donations. However, we suspect that by using an invalid address, scammers compel victims to contact them for assistance, redirecting the communication in their favor. We provide chain analysis on the remaining 78.16\% (111/142) valid addresses to four popular chains:~\emph{Ethereum}, ~\emph{Binance}, ~\emph{Polygon}, ~\emph{Avalance}, and ~\emph{Bitcoin}; identifying 4 of these as suspicious by these popular chains.

\textbf{Incoming Volume/Transfer} In total, we identified 96 accounts with an average USD value of \$2,574,907.09 and a total sum of \$247,191,080.45 at the time of writing this paper. Based on the first transfer date of the transaction, we observe that these 75\% (72/96) were active first in 2024, and the remaining transactions 25\% (24/96) from 2016 to 2013. Scammers using new addresses for transactions are common practices to remain anonymous with the previous transactions history. Among these transactions, we found two long-tail transactions - the first highest recorded account transaction value to \$241,251,535 and the second highest was \$2,863,122.17. Excluding the first and second highest recorded transactions accounts as long-tail, the remaining 94/96 accounts transactions reflected an average of \$32,727.90 and a total sum of \$3,076,422.71 value. Among these 96 transactions, we identified 11 transactions valued less than \$1, with an average incoming volume of \$0.22, and a total sum of \$2.41. We suspect these small incoming transactions below \$1 are rather an airdropping. 

\textbf{Outgoing Transfers} In total, we identified 130 outgoing transfers with an average value of \$1,530.04 and a total sum of \$198,906.

\textbf{Disclaimer.} Our evaluation is based on the observed transaction histories and reported fraud categories. However, are unable to confirm that all transactions associated with these addresses are connected to scams. 

\textbf{}

\begin{mdframed}[style=insightstyle]
\textbf{Key Takeaways.} 
Scammers utilize various payment methods, including PayPal and cryptocurrency, to solicit donations while maintaining anonymity. Our collaboration with industry partners reveals that scammer's payment method linked to various fraud topologies including compliance violations and unauthorized activities. We suspect invalid cryptocurrency addresses are used for manipulating victims to pursue direct communication. The cryptocurrency transaction analysis highlights that scammers often use new addresses to obscure histories, while a small number of accounts perform large sums. Although we could not conclude scams involving transactions of \$1 or less, we suspect that these may go unnoticed due to small recurring payments or platform monitoring biases. Scammers potentially use small transactions, such as airdrops, which may serve to create plausible activity or evade detection.
\end{mdframed}

\section{Dataset Evaluation and Discussion}
\label{sec:data_set_evaluation_and_discussion}

In this section, we provide details on the evaluation of the dataset through manual inspection. We share observed insights into the limitations and assessed the filtration efficacy of using large language models (GPT-4o) and the reliance on external databases for classifying email addresses, phone numbers, and URLs as malicious along with the studied social media profiles.

\textbf{}

\BfPara{Efficacy of LLM-based Filtration} We manually evaluated the effectiveness of using a Large Language Model (LLM) to classify whether a given post is related to a donation context. For this evaluation, we selected 50 posts from each of the social media platforms: \emph{Facebook}, \emph{Instagram}, \emph{Telegram}, \emph{X}, and \emph{YouTube} from both cases, posts that were classified as false and true for donation based context. In total, we manually evaluated 500 posts: 250 from the \emph{True} class and 250 from the \emph{False} class. Our evaluation showed that the LLM achieved 100\% efficacy in correctly identifying donation contexts in the \emph{True} class. However, in the \emph{False} class, we observed two main categories where the LLM underperformed: (i) 19/250 posts lacked sufficient donation contextual information, containing only links, emojis, or hashtags with contact details, and (ii) 33/250 posts found in languages other than English, which were classified as \emph{False}. As a result, we suspect that our evaluation might have overestimated false positive cases while maintaining high true positive efficacy. We propose that these limitations can be further addressed by (i) incorporating additional context checks for prevalent hashtags, and inpsecting the landing URL, and (ii) enhancing the LLM's capabilities to better identify donation contexts in languages other than English through multilingual settings.

\BfPara{Reliability of Security Risk Engines} To assess the reliability of the risk engines used to identify malicious URLs, phone numbers, and emails reported under the abuse category, we conducted two distinct evaluations.

The first evaluation involved inspecting potentially malicious URLs from our dataset by manually opening them in a browser. Out of 252 URLs flagged as phishing or malicious, we randomly selected 100 URLs for inspection. Of these, 47 were inactive or taken down. Among the remaining 53 active URLs, 29 were flagged by \emph{Chrome} as potential phishing or malicious sites with a \emph{Deceptive site ahead} warning. Upon visiting these URLs, we found that 14/29 displayed missing content with a default template, while 13 led to fake donation pages for various causes, such as child support, healthcare, and relief, and 2 were redirected to sign-up pages without further information. For the other 24 active URLs, although they were marked as malicious by the \emph{VirusTotal API}, no deceptive banner was shown upon visiting. However, upon further inspection, each of these 13 URLs was missing content or had been removed, and 11 consisted of solicitations for donations through sign-up or payment information submission pages. For each of these 13 URLs, we found that 1/68 vendors on \emph{VirusTotal} flagged them as suspicious or malicious, while the responses from 68 other vendors were marked as clean. Since phishing sites are often ephemeral and missing content makes classification challenging, not all vendors may have processed these URLs promptly before the content change. We suggest that such cases could be improved through regular monitoring and by consolidating responses from multiple vendors to enhance URL flagging accuracy.

In the second evaluation of phone numbers and email addresses, we conducted additional analyses using two datasets: (i) 50 known malicious entries (25 phone numbers and 25 email addresses) from publicly reported corpus~\cite{scamCorpus}, and (ii) a benign dataset of 50 entries from the authors' friends and family (25 phone numbers and 25 email addresses). We queried these 100 entries against the risk engine and found that 19/25 phone numbers and 23/25 email addresses were flagged with risk levels above 85\%. However, 6 phone numbers and 2 email addresses showed risk percentages between 5\% and 65\%, making them unreliable for classification as malicious. In contrast, all 50 entries from the benign dataset were marked with 0\% risk. Although the risk engine performed inconsistently for email and phone number assessments with lower risk percentages for known corpus, we argue that integrating multiple providers and combining scores could potentially enhance results which would require additional resources.

\BfPara{Social Media Profiles and Scam Prevalence} We randomly selected 100 social media accounts from the dataset and manually inspected them using a browser. Our findings revealed that 9/100 accounts had been deactivated by the social media platforms for violating terms and conditions, and 17/100 were either deactivated or deleted by the users. For the remaining 74/100 active accounts, we manually reviewed their public profiles and engagement. Of these, 14 accounts displayed default profile pictures and had limited public interaction, while 18 accounts were used solely for retweets and shares, with no original posts. We suspect that these accounts are used to harvest followers or create the appearance of an organically aged social media profile. The remaining 42/74 accounts were found to engage in some form of donation solicitation, targeting causes such as ongoing war and human welfare programs (18 accounts), education and local training programs (11 accounts), local wildlife foundations seeking donations for preservation efforts (6 accounts), single mom and women support (3 accounts), and other miscellaneous disadvantaged groups (4 accounts). 
\section{Recommendations}
\label{sec:recommendations}
Based on our observations and findings, we propose recommendations to combat donation-based abuses. These recommendations are intended for adoption by social media platforms, financial services, crowdfunding platforms, and platform users. We provide further details below.

\BfPara{Recommendations to Social Media Platforms} We suggest that social media platforms adopt a detection measurement setup similar to the one proposed in our research. For proactive prevention, social media platforms can utilize a fraud score to assess whether the email address or phone number used during sign-up poses a fraud risk. Similarly, for reactive measures against existing profiles, we recommend monitoring the use of external media associated with profile bio-data or shared posts. Our network analysis of donation abuse revealed that scammers often operate across multiple social media platforms as part of their modus operandi. We encourage social media platforms to share information with other platforms about detected suspicious behaviors to prevent such fraudulent activities. Implementing a warning message for regular users when a social media post contains donation requests from flagged cryptocurrency addresses or payment links could help users avoid potential interactions with scammers.

\BfPara{Recommendations to Financial In-Take Services} We recommend that financial intake services, specifically crowdfunding platforms and payment profiles, monitor the URLs shared across their platforms. For instance, crowdfunding services like \emph{GoFundMe}, \emph{Fundly}, \emph{PayPal}, and others often include links that scammers use to request payouts. These financial intake services can effectively implement referral header monitoring techniques based on the source of visits. Referral headers contain links and source information indicating where a user is directed from. By monitoring referral headers and assessing whether a social media profile is linked to fraudulent activity, crowdfunding platforms, and payment services can reduce the risk of funding abuse by scammers.

\BfPara{Recommendations to Social Media Users} We recommend social media users conduct thorough fact-checking before supporting any donation-related efforts. This includes verifying bio data, and affiliations, understanding the purpose and planned use of funds, and reviewing feedback from other donors. For instance, databases tracking charity affiliations are valuable resources for authenticating charitable organizations. When donating to individuals or private causes, we recommend users support only when there is a known connection and look out for any account duplications or impersonations.

\textbf{}
\begin{mdframed}[style=insightstyle]
\textbf{Key Takeaways.} 
We provide recommendations to combat donation-based abuses on social media platforms, financial services, crowdfunding platforms, and among users. Social media platforms are encouraged to adopt fraud scores for proactive detection and monitor external media for suspicious activity. Financial services are suggested to monitor URLs for scams and use referral header monitoring to reduce fraud risks. Users are urged to conduct thorough checks before donating, verify affiliations, and exercise caution, particularly when supporting unfamiliar causes or individuals.
\end{mdframed}
\section{Conclusion}
In this research, we presented the first large-scale study of donation-based abuses across five social media platforms: ~\emph{X}, ~\emph{Instagram}, ~\emph{Facebook}, ~\emph{Telegram}, and ~\emph{YouTube}. By analyzing data from over 150K social media users and 3 million posts, we identified over 832 scammers soliciting fraudulent donations on these platforms. Our analysis of profile creation and user engagement revealed scammers' techniques for luring victims and requesting payments through payment profiles such as ~\emph{PayPal}, cryptocurrency addresses, crowdfunding services, and survey forms. Our measurement approach identified scam accounts operating on 11 platforms (9 social media, and 2 e-commerce) beyond their origins. Through collaboration with industry partners ~\emph{PayPal} and the cryptocurrency abuse database ~\emph{Chainabuse}, we validated the scams and assessed the financial impact of these fraudulent accounts. Furthermore, we provided detailed disclosures to affected entities and proposed recommendations to protect against future abuses.
\section*{Acknowledgment}
We sincerely thank Ian Schade from Chainabuse for sharing valuable insights regarding cryptocurrency accounts. Our appreciation also goes to Qutub Khan Asghar Vajihi from PayPal for providing insights related to PayPal accounts. Additionally, we are grateful to Muhammad Saad from X (formerly Twitter) for his initial discussions on donation-based scams prevalent in the contexts of the X platform. This work was funded by the German Federal Ministry of Education and Research (BMBF grant 16KIS1900 ``UbiTrans''); and by the EU—NGEU National Sustainable Mobility Center (CN00000023), Italian Ministry of University and Research Decree n. 1033—17/06/2022 (Spoke 10). Lastly, this work was carried out while Dario Lazzaro was enrolled in the Italian National Doctorate on Artificial Intelligence run by the Sapienza University of Rome in collaboration with the University of Genoa.
\bibliographystyle{ieeetr}
\bibliography{strings,bib}

\appendix
\section{Prompt Engineering on Donation Context} 
\label{sec:prompt_engineering_donation_context}
In this section, we provide details on creating prompt injection in identifying the posts that are related to the donation context. We chose LLMs specifically for their effectiveness and adaptability in handling diverse natural language processing tasks, making them ideal for accurately classifying fraudulent donation solicitations. 

To determine if a post is related to donation solicitations, we designed a prompt that evaluates whether the input post includes donation requests, outputting the result as a boolean (true or false). Using the OpenAI API~\cite{openAI}, we queried posts from the five social media platforms to obtain their respective outputs. Below, we provide examples of prompt instruction along with input samples for responses received in both cases (false and true).

\BfPara{Prompt Instruction}
\begin{mdframed}[style=insightstyle]
\begin{quote}
You are given a text and must identify whether it is requesting money, donations, or charity support. The output should be a boolean value compatible with a Python boolean value. Do not include any explanation.
\end{quote}
\end{mdframed}

\BfPara{Input Sample Post - API Response True Case}
\begin{mdframed}[style=insightstyle]
\begin{quote}
WE JUST HIT OUR GOAL OF \$500 of donations to Extra Life. We would like to thank everyone who donated to this great cause!
\end{quote}
\end{mdframed}

\BfPara{Output of ChatGPT - API Response True Case}
\begin{mdframed}[style=insightstyle]
\begin{quote}
True
\end{quote}
\end{mdframed}

\BfPara{Input Sample Post - API Response False Case}
\begin{mdframed}[style=insightstyle]
\begin{quote}
RT @bbby**luve: Oi meus amores! We are only 15 days away from Brazil fanmeeting? Are you ready for that amazing night??
\end{quote}
\end{mdframed}

\BfPara{Output of ChatGPT - API Response False Case}
\begin{mdframed}[style=insightstyle]
\begin{quote}
False
\end{quote}
\end{mdframed}

\begin{figure*}[!htbp]
\centering
\includegraphics[clip, trim=2.5cm 1.5cm 2.5cm 1.1cm,width=0.495\textwidth]{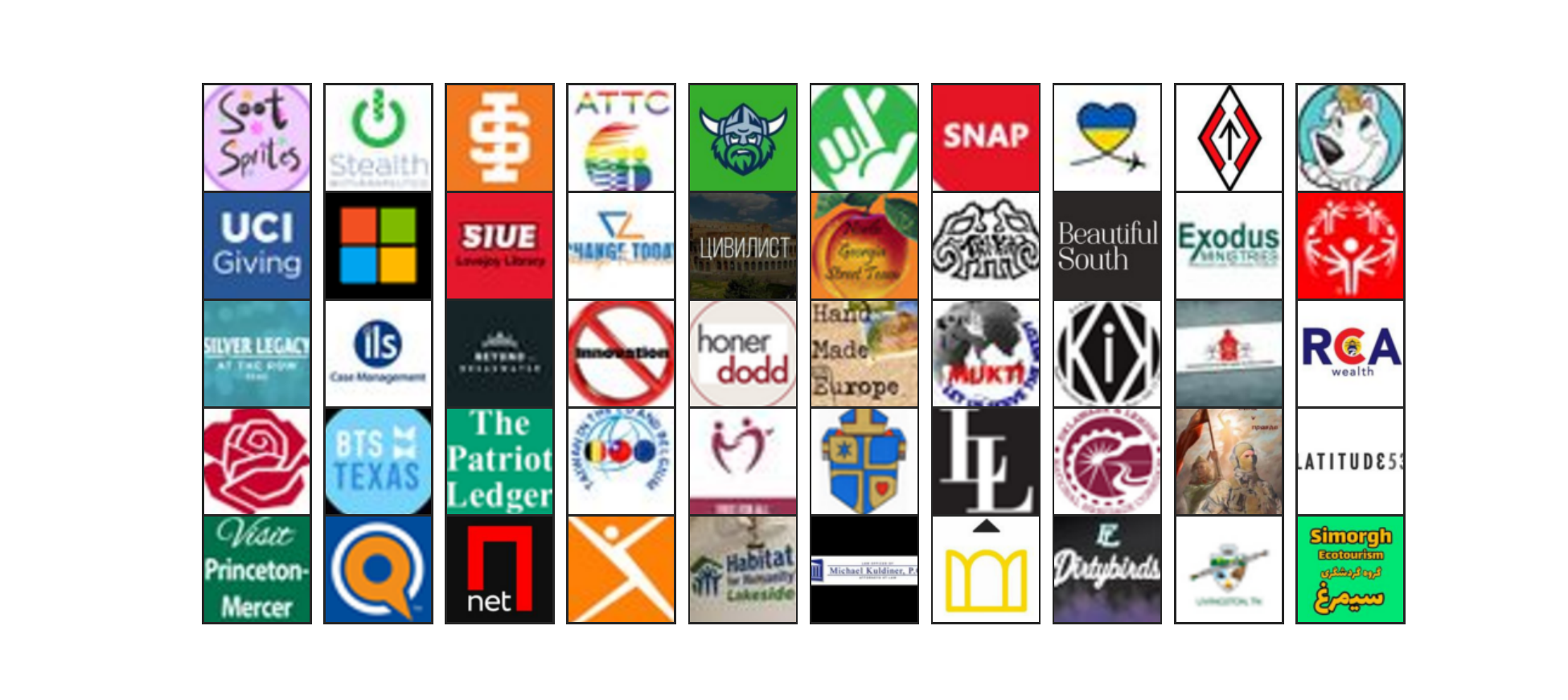}\hfill
\includegraphics[clip, trim=2.5cm 1.5cm 2.5cm 1.1cm,width=0.495\textwidth]{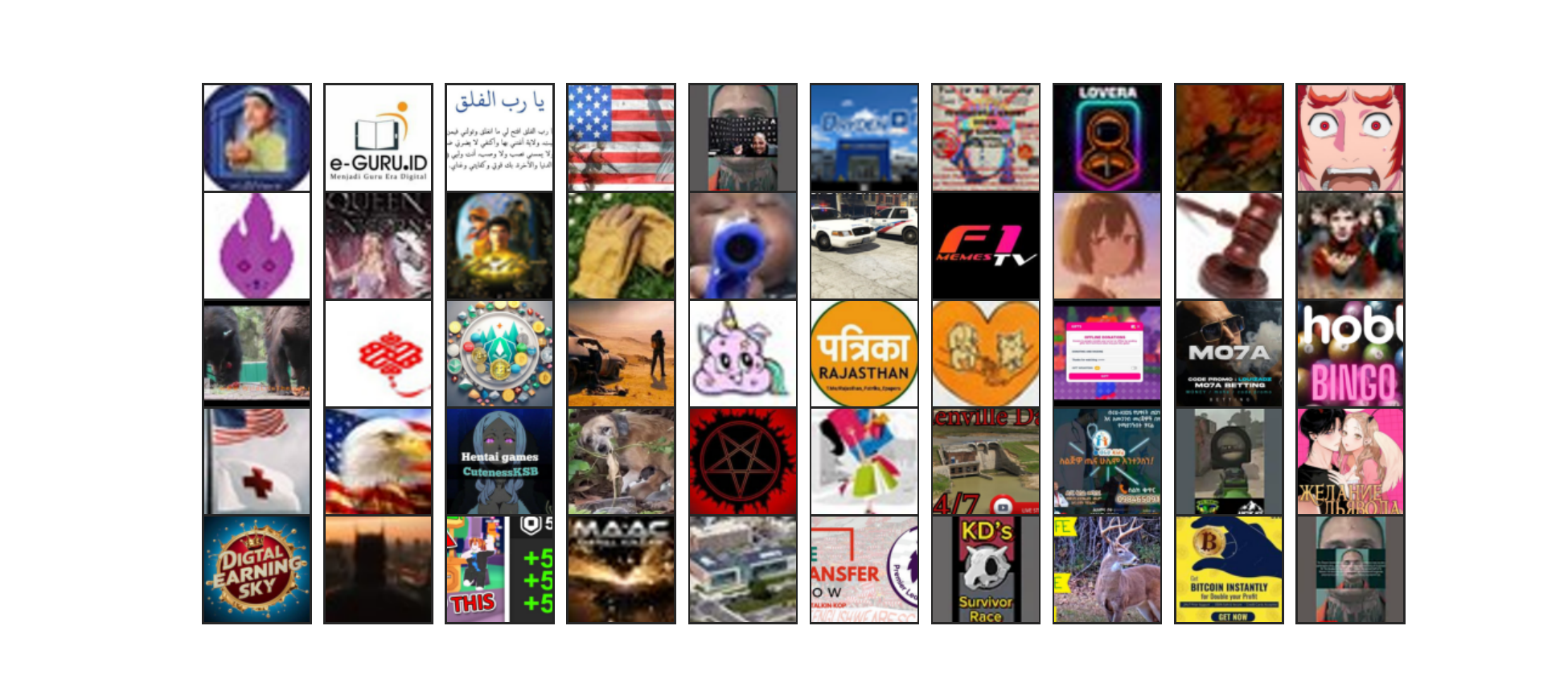}\hfill
\caption{Visualization of 50 random samples from \emph{Association-Logos} (left), \emph{Games-Cartoon} (right) clusters of scammers.}
\label{fig:scammers_cluster_1}
\end{figure*}
\begin{figure*}[!htbp]
\centering
\includegraphics[clip, trim=2.5cm 1.5cm 2.5cm 1.1cm,width=0.495\textwidth]{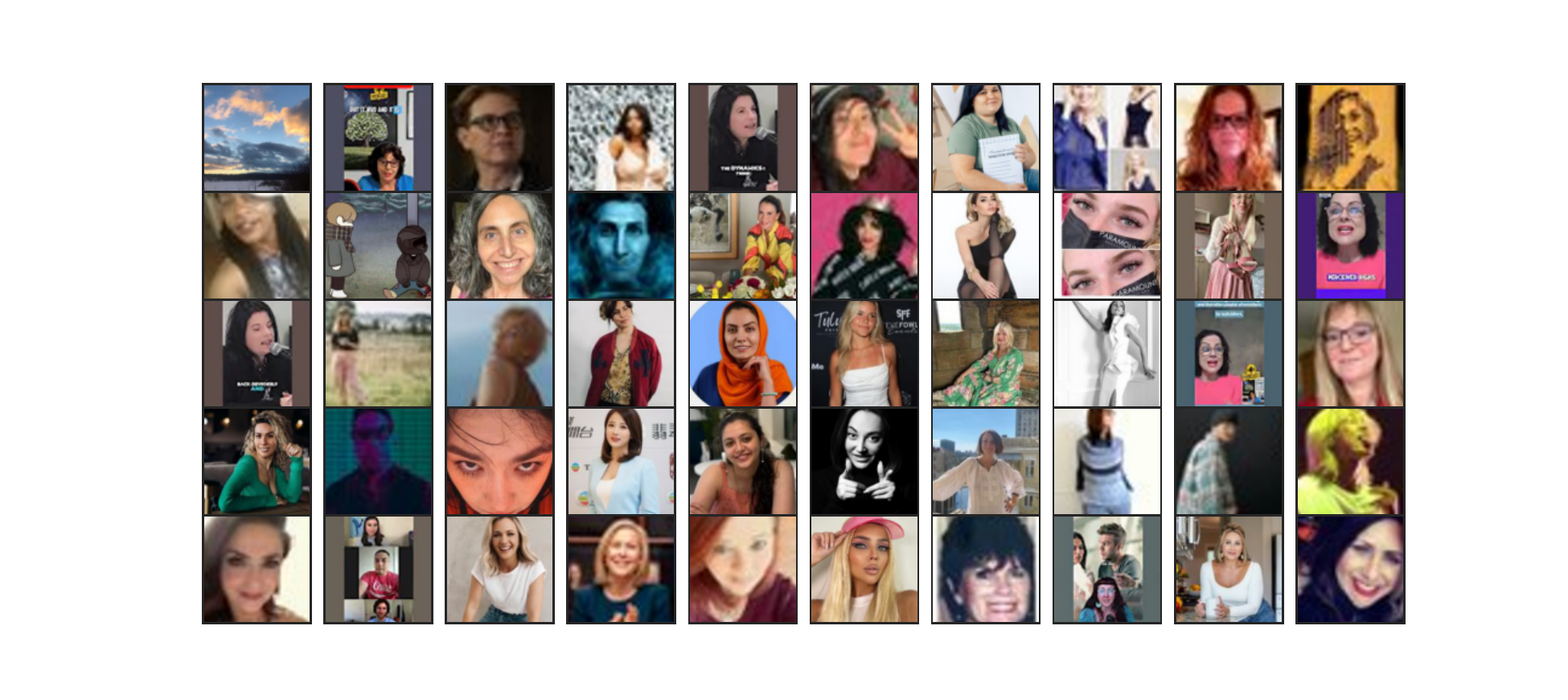}
\includegraphics[clip, trim=2.5cm 1.5cm 2.5cm 1.1cm,width=0.495\textwidth]{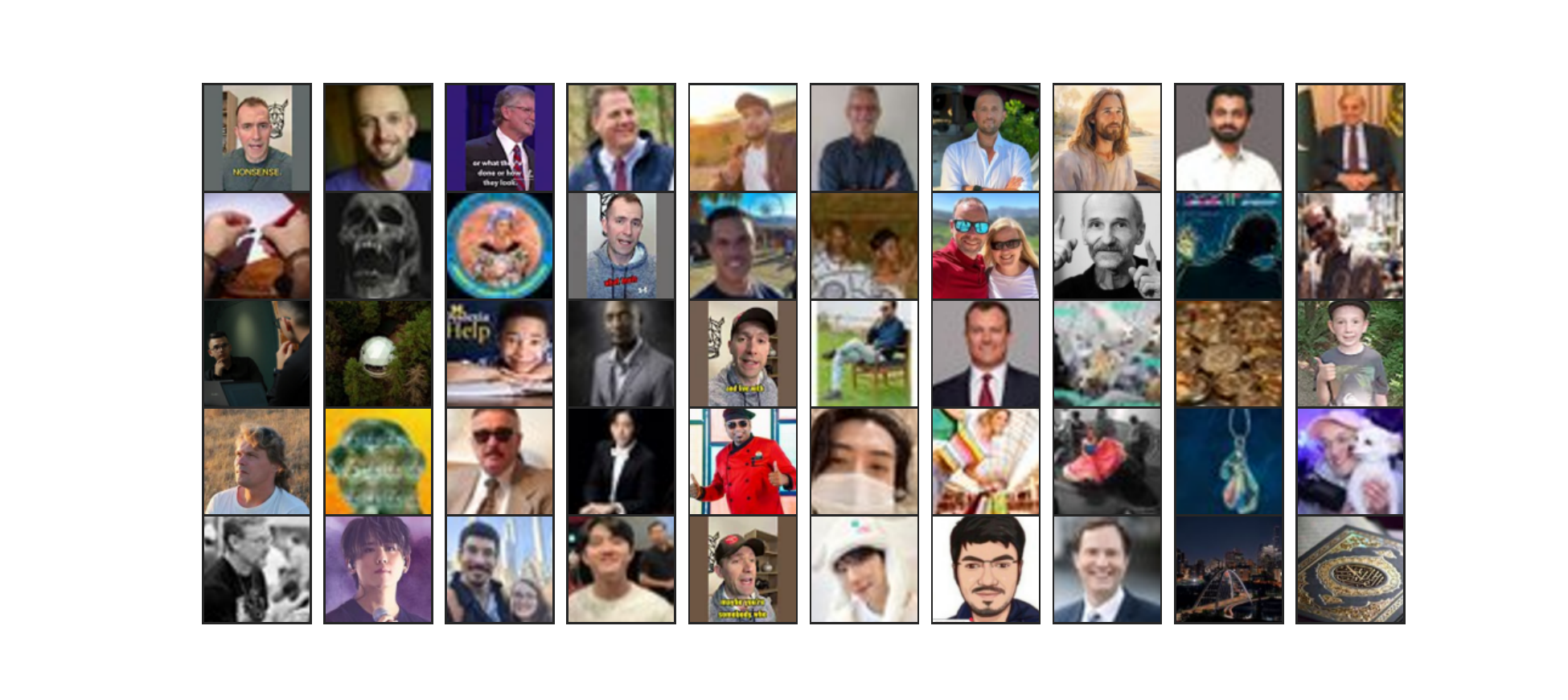}\hfill
\caption{Visualization of 50 random samples from \emph{Female} (left) \emph{Male} (right) clusters of scammers.}
\label{fig:scammers_cluster_2}
\end{figure*}
\begin{figure*}[!htbp]
\centering
\includegraphics[clip, trim=2.5cm 1.5cm 2.5cm 1.1cm,width=0.495\textwidth]{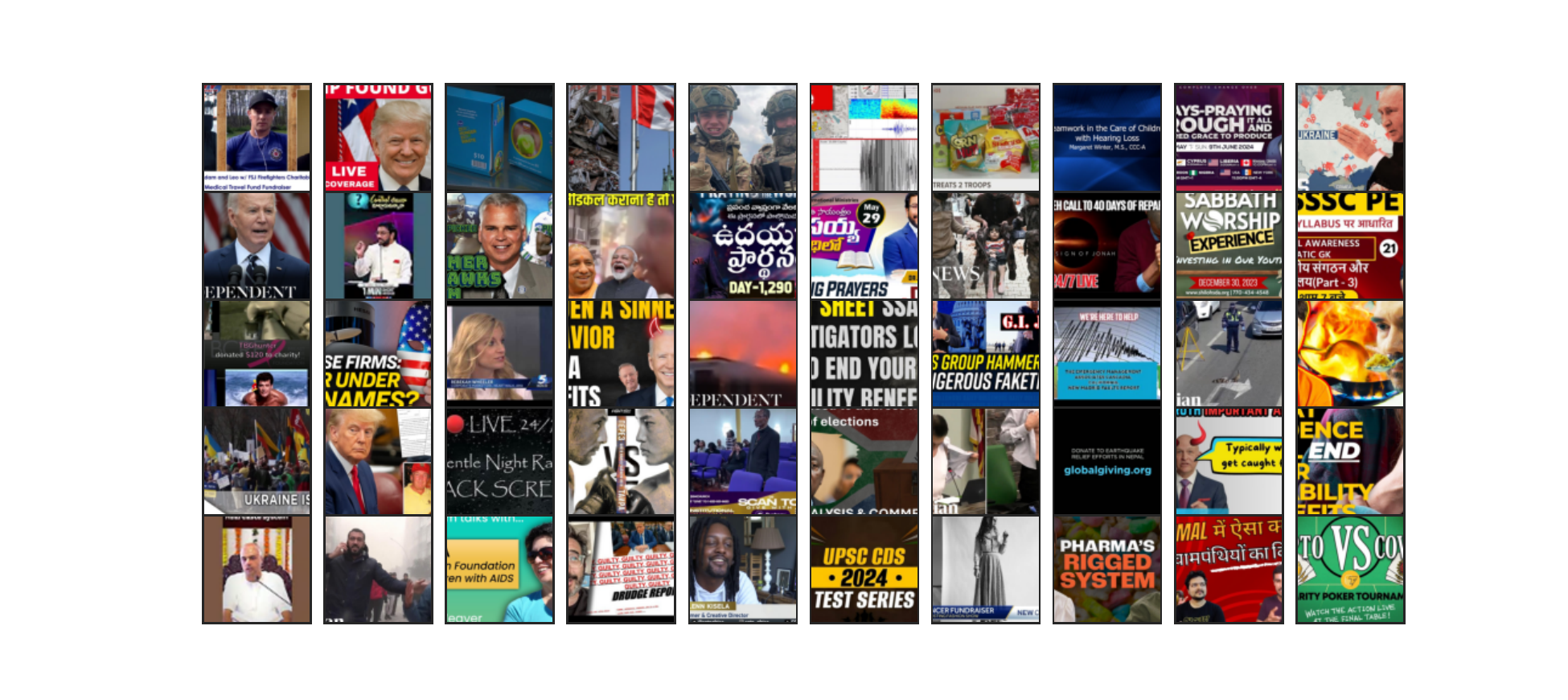}\hfill
\includegraphics[clip, trim=2.5cm 1.5cm 2.5cm 1.1cm,width=0.495\textwidth]{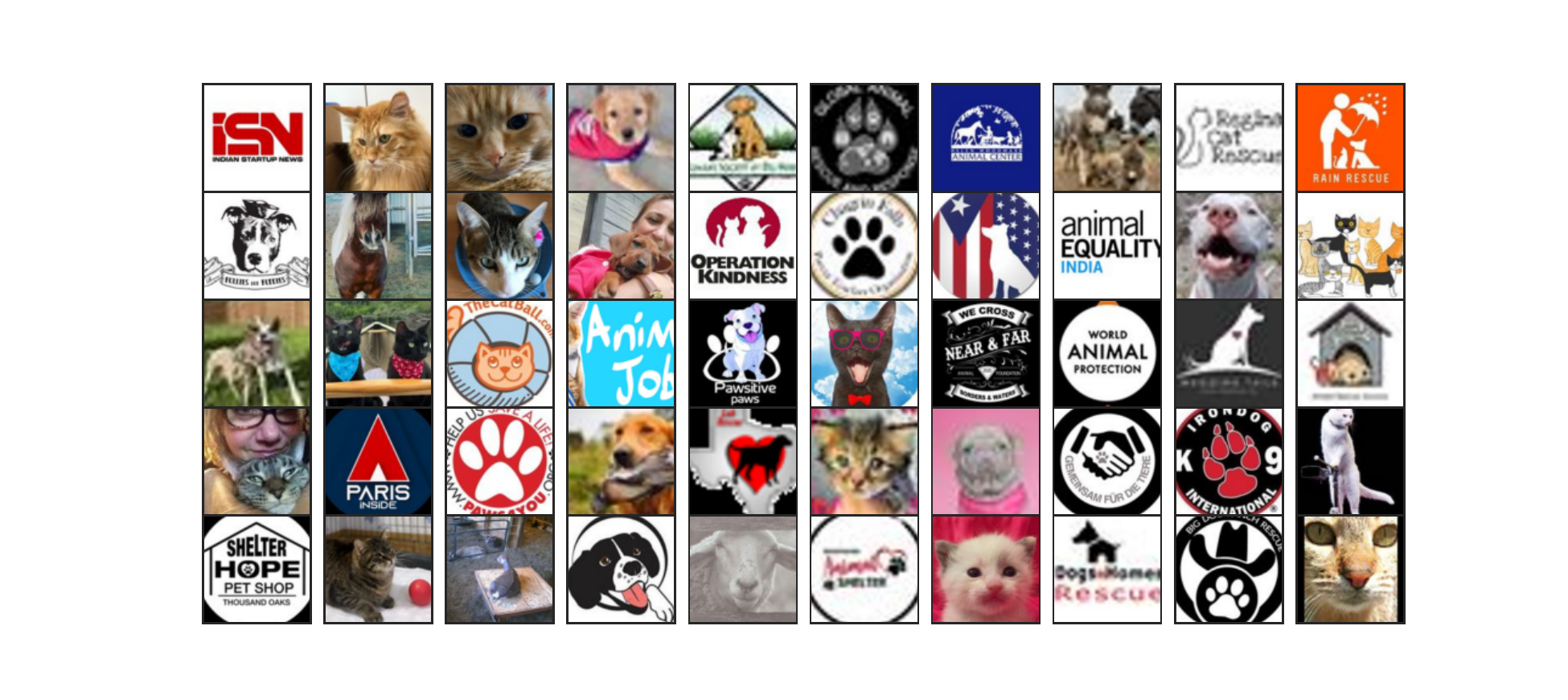}
\caption{Visualization of 50 random samples from \emph{Politics-War} (Left), and \emph{Pets} (right) clusters of scammers.}
\label{fig:scammers_cluster_3}
\end{figure*}
\begin{figure*}[!htbp]
\centering
\includegraphics[clip, trim=2.5cm 1.5cm 2.5cm 1.1cm,width=0.6\textwidth]{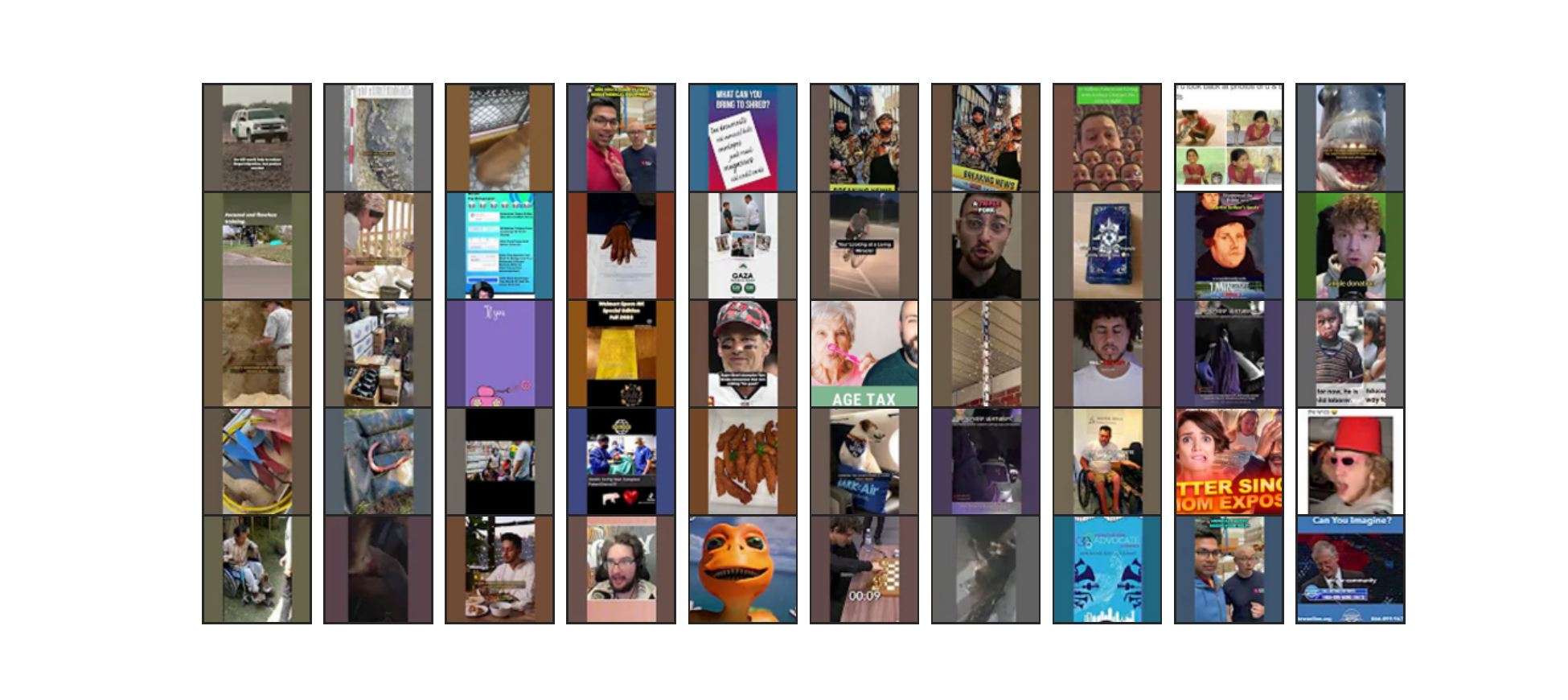}
\caption{Visualization of 50 random samples from Miscellaneous clusters of scammers.}
\label{fig:scammer_cluster_miscellaneous}
\end{figure*}

\end{document}